\documentclass{aa}  

\usepackage{natbib}
\usepackage{graphicx}
\usepackage{rotating,booktabs}
\usepackage[verbose]{placeins}
\usepackage{pdflscape}
\usepackage[utf8]{inputenc}
\usepackage{adjustbox}
\usepackage{tikz}
\usepackage[english]{babel}
\usepackage{graphicx}
\usepackage[amsmath,thref,hyperref]{ntheorem}

\usepackage{txfonts}
\newcommand{\lya}{\ion{Ly}{$\alpha$}}

\newcommand{\ha}{\ion{H}{$\alpha$}}
\defcitealias{Tozzi22}{T22}
\newcommand{\citepgalbiati}{\textcolor{blue}{Galbiati et al. in prep.}}
\newcommand{\citegalbiati}{\textcolor{blue}{Galbiati et al.} (\textcolor{blue}{in prep.})}

\newcommand{\citepannella}{\textcolor{blue}{Pannella et al.} (\textcolor{blue}{in prep.})}

\newcommand{\citepwang}{\textcolor{blue}{Wang et al. in prep.}}

\newcommand{\citepcantalupo}{\textcolor{blue}{Cantalupo et al. in prep.}}

\begin{document} 

   \title{X-ray view of a massive node of the Cosmic Web at $z \sim 3$}
   \subtitle{I. An exceptional overdensity of rapidly accreting SMBHs}

    \authorrunning{A. Travascio et al.}
    \titlerunning{X-ray view of a massive node of the Cosmic Web at $z \simeq 3.25$}   
    
   \author{A. Travascio
          \inst{1}
          \and
          S. Cantalupo
          \inst{1}
          \and 
          P. Tozzi
          \inst{2}
          \and  
          F. Vito
          \inst{3}
          \and 
          A. Paggi
          \inst{4,5,6}
          \and 
          G. Pezzulli
          \inst{7}
          \and 
          M. Elvis
          \inst{8}
          \and 
          G. Fabbiano
          \inst{8}
          \and 
          F. Fiore
          \inst{9}
          \and 
          M. Fossati
          \inst{1}
          \and 
          A. Fresco
          \inst{1}
          \and 
          M. Fumagalli
          \inst{1,9}
          \and 
          M. Galbiati
          \inst{1}
          \and 
          T. Lazeyras
          \inst{1}
          \and 
          N. Ledos
          \inst{1}
          \and 
          M. Pannella
          \inst{9}
          \and 
          A. Pensabene
          \inst{1}
          \and 
          G. Quadri
          \inst{1}
          \and 
          W. Wang
          \inst{1}
          }

\institute{Dipartimento di Fisica “G. Occhialini”, Università degli Studi di Milano-Bicocca, Piazza della Scienza 3, I-20126, Milano, Italy \\ e-mail: \texttt{andrea.travascio@unimib.it} \and 
    INAF - Osservatorio Astrofisco di Arcetri, Largo E. Fermi 5, 50127 Firenze, Italy \and
    INAF - Osservatorio di Astrofisica e Scienza dello Spazio di Bologna, Via Piero Gobetti, 93/3, 40129 Bologna, Italy \and
    Dipartimento di Fisica, Università degli Studi di Torino, via Pietro Giuria 1, I-10125 Torino, Italy \and
    East Asian Observatory, 660 North A'oh$\rm \overline{o}$k$\rm \overline{u}$ Place, Hilo, Hawaii 96720, USA \and
    Istituto Nazionale di Fisica Nucleare, Sezione di Torino, I-10125 Torino, Italy \and
    Kapteyn Astronomical Institute, University of Groningen, Landleven 12, 9747 AD Groningen, The Netherlands \and
    Harvard-Smithsonian Center for Astrophysics, 60 Garden St., Cambridge, MA 02138, USA \and
    INAF - Osservatorio Astronomico di Trieste, Via G. B. Tiepolo 11, 34143 Trieste, Italy}

 \abstract
   {Exploring SMBH population in protoclusters offers valuable insights into how environment affects SMBH growth. However, research on AGN within these areas is still limited by the small number of protoclusters known at high redshift and by the availability of associated deep X-ray observations.}
   {To understand how different environments affect AGN triggering and growth at high redshift, we investigated the X-ray AGN population in the field of the MUSE Quasar Nebula 01 (MQN01) protocluster at $z \simeq 3.25$. This field is known for hosting the largest \lya\ nebula in the \cite{Borisova16} sample, and one of the largest overdensities of UV-continuum selected and sub-mm galaxies found so far at this redshift.}
   {We conducted a ultra deep Chandra X-ray survey (634 ks) observation of the MQN01 field and produced a comparative analyses of the properties of the X-ray AGNs detected in MQN01 against those observed in other selected protoclusters, such as Spiderweb and SSA22.}
  {By combining the X-ray, deep MUSE and ALMA data of the same field, we identified six X-ray AGNs within a volume of 16$\rm ~cMpc^2$ and $\pm 1000 \rm ~km~s^{-1}$, corresponding to an X-ray AGN overdensity of $\approx 1000$. This overdensity increases at the bright end, exceeding what was observed in the Spiderweb and SSA22 within similar volumes. The AGN fraction measured in MQN01 is significantly higher ($f_{AGN} > 20\%$) than in the field and increases with stellar masses, reaching a value of 100$\%$ for $\log(M_*/M_{\odot}) > 10.5$. Lastly, we observe that the average specific accretion rate ($\lambda_\text{sBHAR}$) for SMBH populations in MQN01 is higher than in the field and other protoclusters, generally increasing as one moves toward the center of the overdensity.} 
   {Our results, especially the large fraction of highly accreting SMBHs in the inner regions of the MQN01 overdensity, suggest that protocluster environments offer ideal physical conditions for SMBH triggering and growth.}


   \maketitle

\section{Introduction}

The active phases of galactic nuclei (AGN) represent relatively short periods \citep[$\rm \lesssim 0.1~Myr$;][]{King15,Schawinski15} in the life of a galaxy, during which the central supermassive black hole (SMBH) accretes mass. These brief episodes involve the release of a significant amount of energy, which are expected to influence the evolution of galaxies \citep[][and references therein]{Binney95,Magorrian98,Harrison24} and the surrounding gas on scales extending up to hundreds kpc scales \citep{Tytler95,Fabian03,Cicone15,Travascio20a}. 
The study of the AGN duty cycle, influenced by feeding and feedback processes, is crucial for understanding the role of these phases in galaxy evolution. Recent research by various authors \citep{Bongiorno12,Aird12,Bongiorno16,Aird18,Carraro20} has explored the connection between SMBH accretion and galactic properties, focusing on the effects of AGN feedback on galactic quenching \citep[e.g.,][]{Sturm11,Dubois13,Bongiorno16} and the history of SMBH-galaxy co-evolution \citep{Kormendy13}. This co-evolution is key to explaining the local relationships observed between the masses of galactic bulges and SMBHs  \citep{Haring04}, as well as the similar cosmic evolution trends of SMBH accretion density and star formation rate density \citep{Boyle98,Aird15}.
The role of the environment adds another layer of complexity to this scenario. The AGN phase appears to be particularly significant in high-redshift overdense regions, where abundant gas reservoirs and high merger rates can promote SMBH growth and accelerate galaxy evolution \citep{Martini13,Assef15,Hennawi15,Marchesi23,Vito23,Elford24}.
Studying the AGN activity in these environments is crucial to explore the growth of SMBHs over cosmic time and to achieve a comprehensive understanding of galaxy and large-scale structure evolution, as well as their complex interactions.

Those high-redshift over-densities that are identified as the progenitors, not yet virialized, of the massive ($\gtrsim 10^{14} \rm~M_{\odot}$) galaxy clusters observed in the local Universe, are defined as protoclusters \citep[see][and references therein]{Overzier16}.
Various tracers are employed for identifying protoclusters. High-redshift radio galaxies, exhibiting extensive radio jets, were initially regarded as indicators of potential protoclusters \citep{Carilli97,Pentericci00,Miley08,Hatch14}. Other proxies of protoclusters are significant overdensities of various types of galaxies, e.g., Lyman-$\alpha$ emitters (LAEs), Lyman Break Galaxies (LBGs), and submillimiter galaxies (SMGs).
The presence of large reservoirs of gas is also a key identifier of these structures. The hot phase of diffuse gas can be detected through extended X-ray emission \citep{Gobat11,Stanford12,Tozzi15,Champagne21,Lepore24} and the Sunyaev-Zel'dovich effect in submillimeter data \citep{Gobat19,Andreon23,DiMascolo23,vanMarrewijk23}, and its presence may be associated to large scale shocked gas, or to high density gas in virialized subhalos within the protocluster. The warm phase ($T\sim 10^4-10^5~\rm K$) is revealed by the presence of extended \lya\ emitting structures, typically powered by the photoionization of nearby AGN \citep[e.g.,][]{Cantalupo14,Hennawi15,Battaia18a}. In recent decades, the number of detected protoclusters has been increasing, primarily due to the refinement of selection techniques and the proliferation of surveys dedicated to this purpose \citep{Cucciati14,Toshikawa16,Higuchi19,deCosta21,Uchiyama22}.

Some progress has been made in the search for the AGN population within protoclusters. X-ray observations are crucial for probing AGNs in these environments, as X-ray photons can penetrate high column density gas. This gas causes intrinsic obscuration of softer X-ray photons from the nuclear region and complete obscuration when $\rm N_H \gtrapprox 10^{24}\rm ~cm^{-2}$. Such obscuration is a common characteristic of AGNs in dense regions \citep{Menci03,Hopkins06,Assef15,Vito20}. The X-ray Chandra telescope has played a pivotal role in identifying X-ray AGNs at high redshifts, thanks to its high spatial resolution and sensitivity. Several studies \citep{Lehmer09,DigbyNorth10,Macuga19,Tozzi22,Vito24} conducted with over 200 ks of Chandra observations have provided evidence of enhanced X-ray nuclear activity in protoclusters compared to the field \citep{Georgakakis15,Ranalli16,Vito18}, as well as to lower-redshift galaxy clusters \citep{Martini13}. However, a more in-depth investigation into the effects of these environments on the AGN phase and vice versa requires a greater number of studies based on homogeneous multi-wavelength observations. Such comprehensive data are essential to reach robust conclusions.

This paper is the first contribution in a series dedicated to analyzing Chandra X-ray data of the Multi Unit Spectroscopic Explorer (MUSE) Quasar Nebula 01 (hereafter MQN01) field. This field was originally observed with the MUSE instrument on the Very Large Telescope (VLT) telescope, pointing one of the brightest QSOs in the Universe, CTS G18.01 ($\lambda L_{1700} = 6.5 \times 10^{46}~\rm erg~s^{-1}$). Around it, \cite{Borisova16} found one of the two largest ($> 100 ~\rm kpc$) \lya\ nebulae observed in their sample, potentially correlated with the presence of a protocluster structure. The existence of a protocluster in this field is also supported by numerous multi-wavelength observations acquired over the past three years, all aimed at investigating the galaxy population around this huge \lya\ nebula.
MUSE follow-up observations consisting of a mosaic of four frames, $\sim$10 hr each one, was obtained in a $\sim 4 ~\rm arcmin^2$ area around the QSO CTS G18.01 (PI Cantalupo, 0104.A-0203(A)). These deep MUSE observations reveal a \lya\ emitting structure (\citepcantalupo) more extended than the nebula originally discovered by \cite{Borisova16}.
Moreover, the analysis conducted in \citegalbiati\ revealed 21 galaxies embedded in the protocluster within $\pm 1000~\rm km~s^{-1}$ from the bright QSO CTS G18.01, which was assumed to be the center of the overdensity. This detection indicates an overdensity of $\simeq 53 \pm 17$  of galaxies with a magnitude $\rm M_R <-19.25~\rm mag$.
Finally, the detection of six CO(4-3) emitting galaxies in this field, as probed by ALMA in \cite{Pensabene24}, indicates an overdensity of $\simeq 15 \pm 12$. These high overdensities of spectroscopically-selected protocluster members further confirm the existence of a protocluster in the MQN01 field. 
Additional rest-frame optical and near-UV images obtained with VLT/FORS2, VLT/Hawk-I, HST/F625W, HST/F814W, and JWST/NIRCam will be utilized in forthcoming publications to explore several properties of the galaxy population in the MQN01 protocluster (\citepgalbiati, \citepwang).

In this paper, we present an analysis of deep X-ray Chandra data obtained for the MQN01 field. Our aim is to conduct a comprehensive census of X-ray AGNs, examining their X-ray luminosities, spatial distribution and overdensity within the MQN01 protocluster.
Additionally, we investigate the collective properties of these AGNs, including their fraction relative to galaxies selected based on a stellar mass threshold, and the SMBH accretion rates as a function of environmental characteristics. To ensure an accurate interpretation about the effect of the environment on the AGN activity, these findings are compared with those observed in the field and in other protoclusters.

The Paper is organized as follows. Section~\ref{sec:Chandradata} provides an overview of the Chandra X-ray observations, including a description of the processes of data reduction and of the detection of X-ray sources in the entire Chandra field. In Section~\ref{sec:Xproto}, we present the ancillary multi-wavelength data used to support our selection of the X-ray sources associated with protocluster members. Section~\ref{sec:spectraanalysis} presents the results from the spectral analysis of the X-ray emission from these protocluster members. 
Section~\ref{sec:AGNcompover} focuses on the collective properties of the AGN population in MQN01, including aspects such as compactness, the profile of the hard X-ray Luminosity Function, and the overdensity, all in comparison to the field and the AGN populations in two other protoclusters at $z>2$, namely SSA22 and Spiderweb, which are introduced in Section~\ref{sec:protolucters} and for which deep (>200 ks) Chandra observations have been analyzed in previous work. 
In Section~\ref{sec:envvsSMBH}, we explore potential correlations between galactic and environmental properties and SMBH activity, specifically investigating the variation of the AGN fraction as a function of stellar mass cuts for member galaxies, and the distribution of SMBH accretion rates in MQN01 and the other two protoclusters. The summary and conclusions are detailed in Section~\ref{sec:sumconc}.

Throughout the paper, we adopted a cosmological model with $H_0 = \rm 67.4~km~s^{-1}~Mpc^{-1}$, $\Omega_{\Lambda} = 0.714$, and $\Omega_{m} = 0.286$, consistent with the \cite{Planck18} results. Under this cosmology, 1 arcsecond at $z=3.25$ corresponds to a physical scale of 7.663 kpc. The uncertainties presented in all the plots correspond to a $1 \sigma$ confidence level, unless otherwise specified\footnote{For the X-ray fit parameters, we used a commonly adopted confidence level of 90$\%$}.

\section{Chandra data: reduction and analysis}\label{sec:Chandradata}

\begin{figure}[t]
   \begin{center}
   \includegraphics[height=0.315\textheight,angle=0]{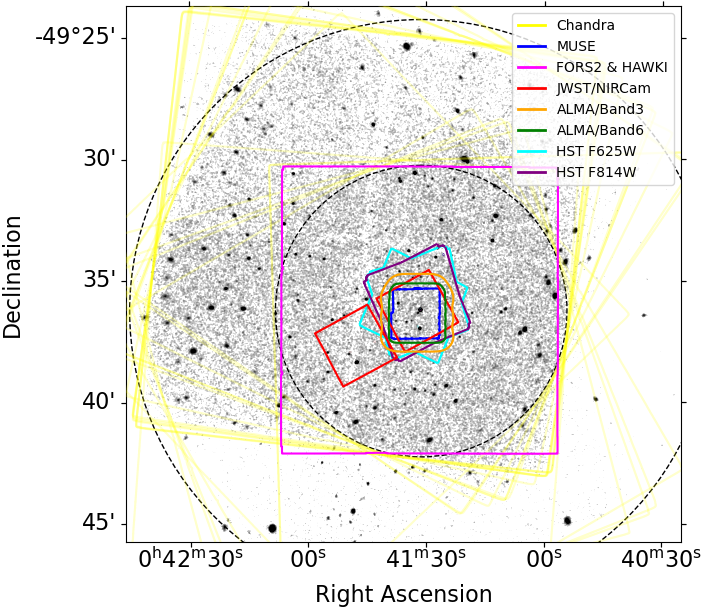}
   \caption{Chandra X-ray image in the  0.5–7.0 keV energy band, smoothed with a $\sim$1.5 arcsec Gaussian filter. Yellow squares mark the individual ObsID of the Chandra observations detailed in Table.~\ref{table:ObsID}. The dashed black lines represent concentric circles centered on the aim point in the final merged image, with radii of 6 and 12 arcminutes. Additional contours delineate the areas of multi-wavelength observations, the details of which are outlined in Section~\ref{sec:multiW}.}
   \label{fig:ChandraField}
   \end{center}
\end{figure}

\subsection{Chandra data}

\setlength{\tabcolsep}{3.3pt}
\begin{table}[!t] 
\caption{\textit{Chandra} observation log.}
\label{table:ObsID}
\begin{tabular}{ccc|ccc}
\hline
\hline
ObsID & $\rm T_{exp}~(ks)$ &  Start Date & ObsID & $\rm T_{exp}~(ks)$ &  Start Date    \\[3pt]
\hline    
25375  & 38.405  & 2023-07-16   &  25721  & 33.098  & 2022-07-09 \\
25711  & 28.711  & 2023-06-13   &  25722  & 28.532  & 2022-11-02  \\
25712  & 30.510  & 2023-07-06   &  25723  & 29.688  & 2022-11-05 \\
25713  & 30.113  & 2023-07-20   &  25724  & 29.338  & 2023-07-24 \\
25714  & 13.906  & 2023-01-21   &  25725  & 29.995  & 2022-10-08 \\
25715  & 34.629  & 2023-06-07   &  25726  & 26.563  & 2023-06-25 \\
25716  & 29.521  & 2023-07-06   &  25727  & 27.053  & 2022-10-29  \\
25717  & 22.794  & 2023-09-28   &  25728  & 40.554  & 2023-08-30  \\
25718  & 32.875  & 2023-06-21   &  25729  & 27.726  & 2023-06-08 \\
25719  & 29.697  & 2023-07-11   &  25730  & 29.508  & 2023-07-11 \\
25720  & 26.734  & 2023-06-29   &  27667  & 13.906  & 2023-01-21 \\
\hline
\hline
\end{tabular}
\end{table}

We analyzed observations taken with the X-ray Chandra Advanced CCD Imaging Spectrometer (ACIS-I), optimized for imaging wide fields, during the Cycle 23 in 2022/2023, with a cumulative exposure time of 634 ks (PI: S. Cantalupo). These observations were conducted in the Very Faint (VFAINT) mode, which enables a better distinction between good and bad X-ray events by utilizing the $5 \times 5$ pixel event island. A list of the ObsID their corresponding pre-data reduction exposure times is provided in Table~\ref{table:ObsID}. Figure~\ref{fig:ChandraField} shows the combined 0.5-7.0 keV image from Chandra observations, smoothed with a $\sim$1.5 arcsec Gaussian filter, with each ObsID observation indicated by a yellow square. We overlay the fields of view of the ancillary data used in this paper to compile a comprehensive census of X-ray AGNs associated with the protocluster members. Further details on these data are provided in section~\ref{sec:multiW}.

\subsection{Data Reduction, Alignment and Merge of the Observations} \label{sec:Chandrareduction}

Data reduction was carried out for each individual Chandra ObsID using CIAO 4.15 and the Chandra Calibration Database (CALDB-4.10.2) installed on Python 3.10. For the level 1 event files, we applied the \texttt{destreak} procedure to eliminate streak events along the pixel rows. Subsequently, we employed \texttt{acis\_build\_badpix} and \texttt{acis\_find\_afterglow} to generate a bad pixel file and identify events potentially associated with residuals from cosmic rays. Given that the observations were conducted in VFAINT mode, we set the parameter \texttt{check\_vf\_pha}=yes for a more accurate background cleaning. 
The event files were then filtered for standard event grades of 0, 2, 3, 4, and 6. A visual inspection was conducted on these event files to identify hot pixels and assess pile-up effects. Then the \texttt{acis\_process\_events} tool was used to update the data and correct for the effects of temperature-dependent charge-transfer inefficiency (CTI).  
We run \texttt{wavdetect} on each chip in the individual ObsID, setting the scale parameter to 2, 4, and 8, and the detection threshold to 1e-07. We used \texttt{wavdetect} output to identify the positions of X-ray detections, which were subsequently removed from the event files to generate new event files containing only background emission. From these files, we extracted light curves with a time binning of 100 s. Hence, we utilized \texttt{deflare} to filter the original event files for time intervals in which the background exceeds the average value by $\rm 3 \sigma$, after applying a clipping algorithm. Finally, all the chips were merged using the \texttt{dmmerge} tool. Following the filtering of dead-time due to flares identified in light curves we obtained a final dataset with exposure time of $\sim$634 ks. Finally, we created the aspect solution files for each CCD using \texttt{asphist}, and we utilized the \texttt{mkinstmap} and \texttt{mkexpmap} tools on individual chips and ObsID to generate monochromatic exposure maps at 1.5 and 4.5 keV.

To align the astrometry of each ObsID, we used \texttt{flux\_obs} and \texttt{mkpsfmap} tools, to create the broad energy band (0.5-7.0 keV) images and the Point Spread Function (PSF) maps at the energy 2.3 keV, respectively. The PSF map was weighted according to exposure times (i.e., \texttt{psfmerge="exptime"}), with each pixel's value representing the PSF radius containing an Encircled Counts Fraction (ECF) of 0.393 (i.e., \texttt{psfecf = 0.393}).
These products were used to identify X-ray detections with \texttt{wavdetect}, setting the wavelet scale following the $\sqrt{2}-$sigma series\footnote{Corresponding to the wavelet scales ''1.0 1.414 2.0 2.828 4.0 5.657 8.0 11.314 16.0''.} and specifying that we expect one false source rate per scale by setting \texttt{falsesrc=1}. We used the positions of the sources in common between different observations, using as reference the deepest ObsID 25728, to reproject the event files, the monochromatic exposure maps weighted by time (unit=$\rm cm^2$), and the aspect solution files, using the \texttt{wcs\_match} and \texttt{wcs\_update} tools. This allowed us to align the different ObsID that will be merged. Subsequently, we used the GAIA catalog to correct the astrometry of the merged event file, initially showing a shift with respect to the GAIA sources of $\approx 0.55''$.

After aligning the aspect files and performing the astrometric correction, we combined the monochromatic exposure maps using the corresponding exposure times as weights, and a merged event file from all the ObsIDs was then obtained by applying \texttt{reproject\_obs}. Figure~\ref{fig:ChandraField} displays the smoothed image with a 3-pixel kernel radius in the 0.5-7.0 keV energy band of the merged event file.
The aim point of the merged event file was determined from the monochromatic exposure map at 1.5 keV as the position of the peak value ($\rm RA_{aim~point}$ = 10.3801543 deg, $\rm DEC_{aim~point}$ = -49.6043256 deg). Figure~\ref{fig:ChandraField} displays two circles centered at the aim point, with radii of 6 and 12 arcmin. A radius of 6 arcmin is equivalent to $\rm \approx 2.8~pMpc$ i.e., 11.9 cMpc at $z=3.25$. 

\subsection{X-ray sources detection} \label{sec:xdetections}

Our main objective is to obtain a catalog of X-ray sources in the $\rm 4 ~arcmin^2$ region covered by the MUSE FoV, centered on the aim point of the Chandra observations, because only in that region we have spectroscopically confirmed protocluster members.
We used \texttt{wavdetect} to produce a complete catalog of Chandra X-ray sources in this specific region, minimizing false detections and missed detections.
The \texttt{wavdetect} tool was run by setting the false-positive probability threshold to $10^{-5}$ (i.e., \texttt{falsesrc=-1} and \texttt{sigthresh=1e-5}), and the scale parameter to \texttt{scales = "1 1.41 2 2.83 4 5.66 8 11.31 16"}. Using this value for the threshold \texttt{sigthresh}, we expected $\approx 17$ detections of background fluctuations within 6 arcmin radius from the aim point.
This process results in X-ray point-source catalogs, with 210 X-ray sources in the soft energy band, 313 in the broad band, and 275 in the hard band. We combined these three catalogs to generate a master catalog comprising 360 X-ray point-sources detected in at least one energy band.
A more detailed selection using a signal-to-noise (S/N) threshold identified 311 sources with S/N$>$2 and 213 sources with S/N$>$3. However, this selection does not affect our results, which focus on the MUSE mosaic FoV near the aim point. This more detailed analysis will be presented in future studies, aiming at investigating the X-ray counterparts of galaxies in a wider area than that of the MUSE mosaic FoV.
\subsection{Sky Coverage and Number of Counts} \label{sec:convfact}

Given that one of the objective of this paper is to explore the density of any X-ray AGN population in the protocluster MQN01, it is necessary to determine the sky coverage ($\Omega$) of these Chandra observations i.e., the area covered at a given sensitivity.
The first step to obtain the sky coverage curve consists in estimating the conversion factors from counts rate to fluxes, in different energy bands.
The conversion factors were computed by generating ancillary response files (ARFs) and redistribution matrix files (RMFs) at the aim point using the \texttt{mkwarf} and \texttt{mkrmf} tools, respectively, for individual ObsID event files.
ARFs and RMFs data were used to simulate one hundred high-S/N spectra of an X-ray source positioned at the aim point using the \texttt{fake\_pha} tool in \texttt{Sherpa} \citep{Freeman01}, a software package integrated within CIAO and operated via Python. The spectral emission is assumed to follow a power-law distribution with a photon index ($\Gamma$) ranging between $1.2$ and $1.9$, encompassing both absorbed ($\Gamma < 1.7$) and unabsorbed ($\Gamma > 1.7$) AGN \citep{Lanzuisi13,Liu17}. 
Therefore, for each ObsID, the conversion factor at the aim point within a specific energy band was determined as the ratio between the integrated flux (in $\rm erg~cm^{-2}~s^{-1}$) and the net count rate (in $\rm counts~s^{-1}$). These values were obtained from the spectrum utilizing the tools \texttt{calc\_energy\_flux} and \texttt{calc\_model\_sum}.
The final conversion factors and their associated errors, that are reported in Table~\ref{table:convfact}, were computed as the average and standard deviation, respectively, across the conversion factors obtained for different $\Gamma$, ObsIDs and energy ranges.

\setlength{\tabcolsep}{12pt}
\begin{table}[!t] 
\caption{Conversion factors were calculated at the aimpoint and were used to convert any count rate in the energy band listed in the first column to a photometric flux in the energy band listed in the second column.}
\label{table:convfact}
\begin{tabular}{ccc}
\hline
\hline
Count rate & Energy & Conversion Factor \\
band [keV] & band [keV] & [$\rm 10^{-11} ~erg~cm^{-2}~counts^{-1}$] \\[3pt]
\hline     
0.5-2.0  & 0.5-2.0  & 1.790 $\pm$ 0.083 \\
2.0-7.0  & 2.0-7.0  & 2.196 $\pm$ 0.033 \\
2.0-7.0  & 2.0-10.0 & 3.003 $\pm$ 0.113 \\
\hline
\hline
\end{tabular}
\end{table}

\begin{figure}[t]
   \begin{center}
   \includegraphics[height=0.26\textheight,angle=0]{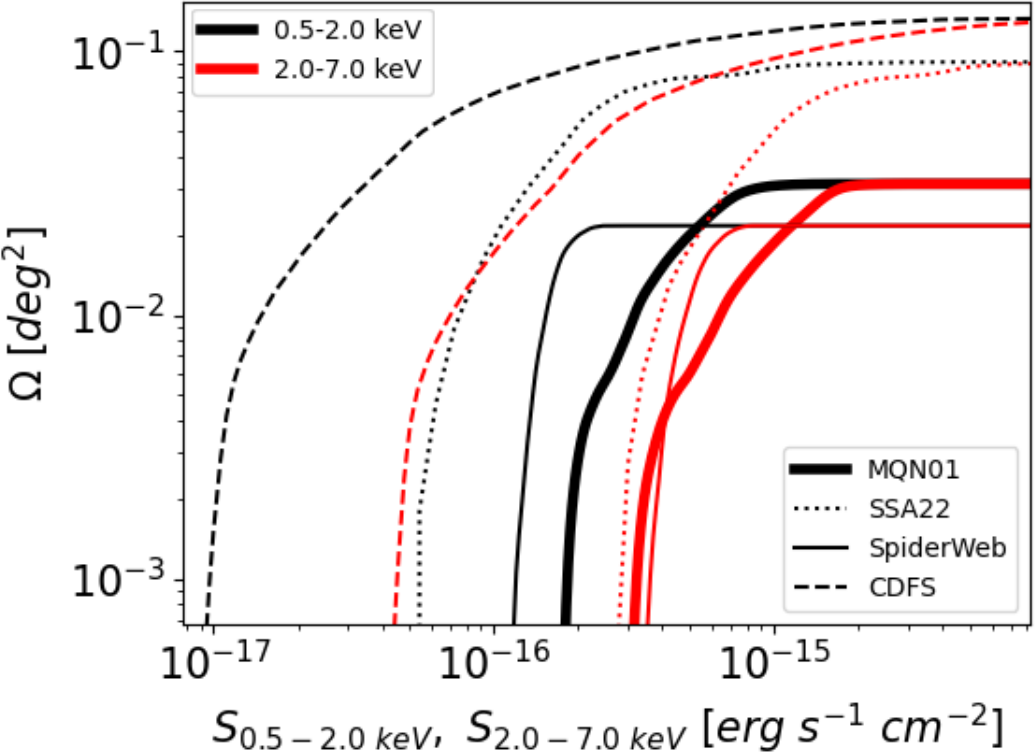}
   \caption{Point source sky coverage for different Chandra observations in the soft (0.5-2.0 keV; black) and hard (2.0-7.0 keV; red) energy band, respectively. The thick solid lines mark the sky coverage of this study, while the thin solid, dotted, and dashed lines depict the sky coverage in the Chandra observations of Spiderweb \citep{Tozzi22}, SSA22 \citep{Lehmer09}, and Deep Field South \citep{Rosati02,Luo17}.}
   \label{fig:SK}
   \end{center}
\end{figure}

\begin{figure}[t]
   \begin{center}
   \includegraphics[height=0.18\textheight,angle=0]{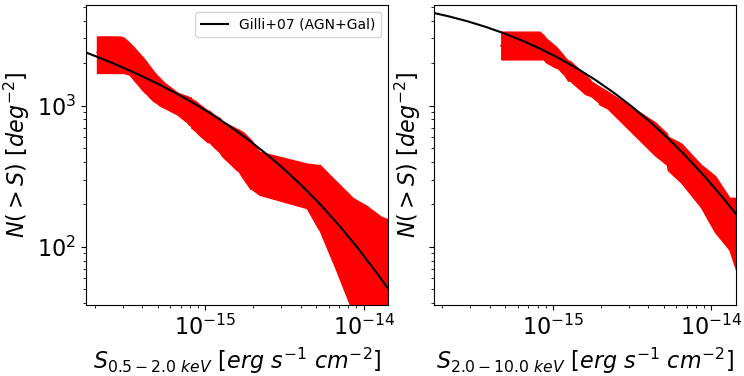}
   \caption{Cumulative number of counts in the soft (left) and hard (2-10 keV; right) energy band, derived within a distance of 6 arcmin from the aim point in the Chandra data of MQN01. The red shaded area shows the uncertainties within 1$\sigma$ due to photometric errors. The black line shows the number of counts predicted for the field from the AGN+Galaxy models developed in \cite{Gilli07}.}
   \label{fig:NS}
   \end{center}
\end{figure}

To generate the curves of sky coverage (black lines in Figure~\ref{fig:SK}), we adopted a completeness criterion of $S/N>2$, considering the observational solid angle within which ideal X-ray sources are detected above a specific flux.
The S/N was calculated based on the net counts estimated in each region of the Chandra observation assuming a specific flux. 
We multiplied the flux of each X-ray source by the exposure time and the effective area value, observed in the 1.5 and 4.5 keV monochromatic exposure maps, at that position. The effective area was normalized to the value observed at the aim point. This product was then divided by the conversion factor, yielding the expected net counts associated to a ideal X-ray source with a specific flux. Hence, for each specific flux, we quantified the associated solid angle according to the number of X-ray sources candidates observed with S/N$>$2, based on the local background.
In Figure~\ref{fig:SK} we report the sky coverage curves for the CDFS \citep[dashed lines;][7 Ms]{Rosati02,Luo17}, which is the field with the deepest X-ray data collected to date, and for other deep ($\rm t_{exp} > 400 ~ks$) Chandra observations obtained for two other protoclusters: SSA22 \citep[dotted lines;][400 ks]{Lehmer09} and Spiderweb \citep[solid lines;][700 ks]{Tozzi22}. 
These estimates encompass areas with varying solid angles and exposure times. The black and red curves represent the sky coverage for soft (0.5-2.0 keV) and hard (2.0-7.0 keV) fluxes, respectively.
The soft and hard flux limits of our observations, defined as those observed within a minimum solid angle $\Omega_{\rm min} = 2 \times 10^{-3} ~\rm deg^2$ equivalent to one tenth of the full FoV, are $1.7$ and $2.9 \times 10^{-16}\rm ~ erg~s^{-1}~cm^{-2}$, respectively. The hard energy flux limits of the observations in all the protoclusters are very similar to each other, about one order of magnitude higher than those in the CDFS observations. In the soft band, the flux limit of our observations is about 3 and 1.5 times higher than those in Spiderweb and SSA22 observations, respectively, and $\sim$20 times larger than those in the 7 Ms CDFS observations.
These differences are due to the combination of the different exposure times of the observations in the various fields, and the decreasing Chandra's effective area in the soft band with time.

Figure~\ref{fig:NS} displays the cumulative number counts $N(>S)$ in the 0.5-2.0 (left) and 2.0-10.0 keV (right) energy bands, of detected X-ray sources in our Chandra observations within 6 arcminutes from the aim point, derived as in \citealt{Rosati02} \citep[see also][]{Luo17,Tozzi22}. The shaded areas show the $2 \sigma$ uncertainties derived by propagating the photometric flux errors of the sources.
The black lines are the prediction from the AGN+Galaxy model developed in \cite{Gilli07} using the "X-ray AGN number counts calculator" in the CXB tool\footnote{http://cxb.oas.inaf.it/counts.html} by taking into account the decline of the AGN number density as a function of redshift. 
We noticed that the presence of a protocluster does not impact the number counts. Hence, the AGN enhancement that we identified, as will be shown in the following sections, lacks significance in amplitude to identify the protocluster as an excess in X-ray counts onto the X-ray sky due to the low statistics. This finding aligns with the results of previous studies analyzing Chandra data in SSA22 and Spiderweb protoclusters \citep{Lehmer09,Tozzi22}.

\section{Probing X-ray Emission in the MQN01 Protocluster: A Comprehensive Data Overview and Cross-Matching Strategy}\label{sec:Xproto}

We present an analysis aimed at associating Chandra X-ray sources with spectroscopically-selected protocluster members, as identified in previous \citep{Pensabene24} and ongoing studies (\citepgalbiati).

\subsection{Ancillary data}\label{sec:multiW}

In recent years, the MQN01 field has been observed using various telescopes and instruments across a wide range of wavelengths. 
Our analysis of X-ray Chandra data will leverage on the catalog of protocluster members spectroscopically selected through ALMA and MUSE observations. Additionally, we will consider galactic properties, such as stellar mass ($M_*$) and star formation rate ($\text{SFR}$), derived through Spectral Energy Distribution (SED) fitting analysis using all available photometric flux measurements. Detailed descriptions of these observations and the respective analyses are provided below.

\subsubsection{Spectroscopic Data: VLT/MUSE and ALMA}\label{sec:MUSE}

VLT/MUSE-WFM Adaptive Optics (AO) observations are obtained from the Guaranteed Time Observation (GTO) program (PI S. Cantalupo), amounting to a cumulative observation time of 40 hours. These observations consist of a mosaic of four pointings, each with a duration of 10 hours, and covering a FoV of $\sim 2 \times 2~\rm arcmin^2$ shown with a blue contour in Figure~\ref{fig:ChandraField}.
\citegalbiati\ identifies spectroscopically selected protocluster members in the MUSE FoV, based on the presence of multiple high-S/N emission and/or absorption lines. They found 21 galaxies within $\pm 1000 \rm ~km~s^{-1}$ of the redshift of the QSO CTS G18.01, which was excluded from their analysis. 

Contours in orange and green in Figure~\ref{fig:ChandraField} delineate the FoV of ALMA 12-m array data, which were obtained in Band 3, centered at $2.94~\rm mm$, and Band 6, centered at $1.26~\rm mm$, during Cycle 8 (Program ID: 2021.1.00793.S, PI: S. Cantalupo), respectively. These data have been analyzed in \cite{Pensabene24}, revealing the presence of six CO(4–3) emitters within $|\delta v_{QSO}| < 1000~\rm km~s^{-1}$.

\subsubsection{Photometric observations: VLT/FORS2, HST, JWST/NIRCam and VLT/HAWK-I}

The magenta square in Figure~\ref{fig:ChandraField} marks the FoV of images taken with the Focal Reducer and low-dispersion Spectrograph 2 mounted on the Very Large Telescope (VLT/FORS2), in U ($\lambda_{cen} = 361~nm$), B ($\lambda_{cen} = 440~nm$) and R ($\lambda_{cen} = 655~nm$) filters with on-source exposure times of $\sim$7~hrs, 5~hrs and 0.75~hrs, respectively. 
In the same area where FORS2 data were collected, HAWK-I images were obtained (programme ID 110.23ZX, PI S.Cantalupo) using CH4 ($\lambda _{eff} = 1.575 ~\rm \mu m$), H ($\lambda _{eff} = 1.620~\rm \mu m$), and Ks ($\lambda _{eff} = 2.146~\rm \mu m$) filters, each with exposure times of 1 hour, 2 hours, and 2 hours on source, respectively.
Furthermore, observations covering $2 \times 5~\rm arcmin^2$ (depicted by red squares in Figure~\ref{fig:ChandraField}) were conducted with the James Webb Space Telescope (JWST) using the extra-wide filters F150W2 ($0.6 - 2.3~\rm \mu m$; pixel scale of 0.031 arcsec/pixel) and F322W2 ($2.4 - 5.0~\rm \mu m$; pixel scale of 0.063 arcsec/pixel). These observations were acquired under GO Cycle 1 program ID 1835 (PI S. Cantalupo), with a single pointing for each filter and an exposure time of 27 minutes.
Additionally, HST/ACS observations (PI: S. Cantalupo, PID: 17065) were taken, during cycle 30, with the ACS F625W and F814W filters, depicted by cyan and purple squares in Figure~\ref{fig:ChandraField}, respectively.

\citegalbiati\ conducted a broad photometric selection of candidates at $z=3-3.5$ using the FORS UBR-selection criterion based on BPASS models plus IGM transmission, as described in \cite{Steidel18}. This selection identified 30 potential protocluster candidates in the MUSE FoV, out of which 12 have confirmed spectroscopic redshifts obtained from VLT/MUSE observations. 
All the photometric data were used, along with Chandra, MUSE, and ALMA data, to perform SED-fitting analysis. Specifically, HAWK-I data are very important for constraining the stellar masses of the galaxies embedded in the protocluster and for selecting Balmer-Break galaxies. High-resolution HST ($\rm FWHM= 0.12''$) and JWST ($\rm FWHM \sim 0.045-0.1''$) images will be crucial for exploring the morphology of galaxies and correlating them with the large-scale environment.

\begin{figure*}[t]
   \begin{center}
   \includegraphics[height=0.38\textheight,angle=0]{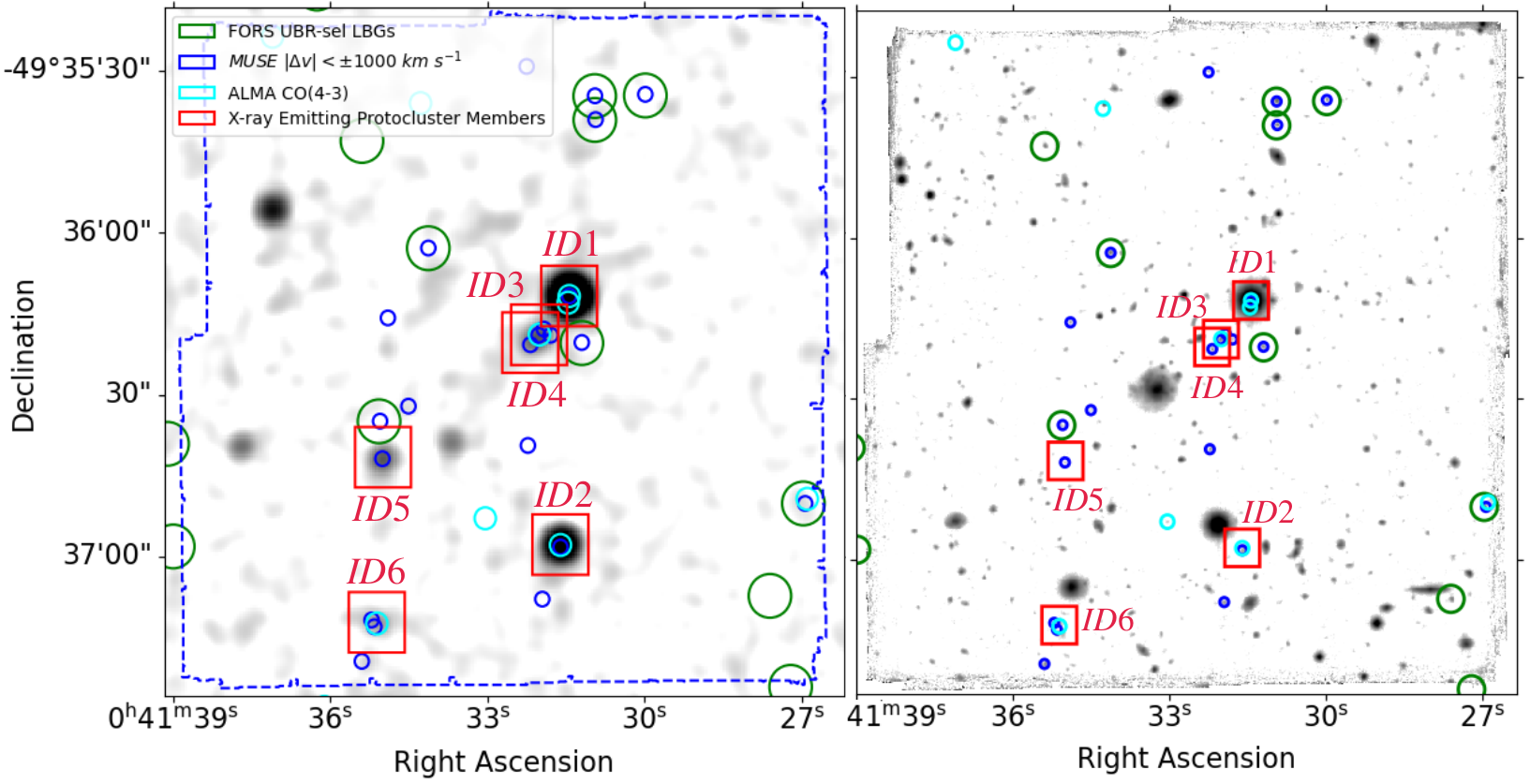}
   \caption{Chandra X-ray and MUSE images of the MUSE field with symbols marking Chandra X-ray point sources and MQN01 protocluster membership candidates identified through spectroscopic and photometric methods. On the left, a 2-pixel Gaussian kernel-convolved image at the energy band of 0.5–10 keV, zoomed into the MUSE FoV (blue dashed square and blue square in Figure~\ref{fig:ChandraField}). Red squares mark the X-ray point sources detected with S/N$\geq$3, each associated with an identity number. Blue and cyan circles show the spectroscopically confirmed MQN01 protocluster members within $\pm 1500\rm~ km~s^{-1}$ from the QSO CTS G18.01 (ID1) from MUSE (\citepgalbiati) and within $\pm 4000\rm~ km~s^{-1}$ from ALMA \citep{Pensabene24}, respectively. The green circles indicate the positions of FORS photometrically selected protocluster member candidates (\citepgalbiati). The right panel shows the MUSE White Light Image of the same field with the same markers for multi-wavelength selections.}
   \label{fig:Detections}
   \end{center}
\end{figure*}

\subsection{Cross-Matching Chandra Data with Multi-Wavelength Datasets}\label{sec:crossmatch}

The primary aim of this section is to compile a comprehensive census of X-ray sources associated with galaxies embedded into the MQN01 protocluster. The redshift of the protocluster is set to that of the QSO CTS G18.01, which is $z_{\text{MQN01}}=3.2502$ as determined by the CO(4-3) line detected in ALMA data \citep{Pensabene24}.
To achieve this objective, we conducted a cross-match between the X-ray source catalog, detailed in Section~\ref{sec:xdetections}, and the spectroscopically selected protocluster members detected with MUSE and ALMA in their FoV.
The X-ray emission is associated with a specific galaxy if the centroid offset is less than 0.5 arcseconds. This threshold is similar to the Chandra PSF (FWHM$\approx$ 0.5 arcsec) and smaller than that of WFM-AO MUSE PSF (FWHM$\approx$ 0.8 arcsec). Using this approach we found six X-ray sources associated to galaxies embedded into the MQN01 protoclusters. 

Figure~\ref{fig:Detections} presents the 2-pixel Gaussian kernel-convolved Chandra X-ray image at 0.2-10 keV (left panel) and the MUSE white-light image (WLI) (right panel). The blue dashed contours in the left panel outlines the FoV of the MUSE mosaic shown on the right. In both panels, spectroscopically selected protocluster members identified with MUSE and ALMA are marked with cyan and blue circles, while red squares indicate the six X-ray sources associated with protocluster members, each labeled with an ID number (ID1-6). 

ID1 corresponds to the bright QSO CTS G18.01. ID2 is the second brightest X-ray emission detected in this area. The remaining X-ray sources are numbered sequentially, with ID3 representing the northernmost source and ID6 representing the southernmost one. 
ID3 and ID4 are two closely positioned X-ray sources ($\sim 2~\rm arcsec$) that can be easily de-blended, partly due to their different energy flux distributions. ID5 is associated to a compact galaxy. ID6 is associated with a very large disc galaxy observed with JWST/NIRCam data and confirmed as a protocluster member based on the spectroscopic redshift revealed by JWST/NIRSpec slit spectroscopic data and ALMA (\citepwang). 
A zoom-in on these X-ray emissions at different energy bands, along with the circles marking the galaxies selected as protocluster members, is presented in Figure~\ref{fig:zoomcounts} in appendix~\ref{app:zoomin}. 
All six X-ray detections are associated with galaxies that were spectroscopically-selected with MUSE and ALMA, within a velocity range of $\pm 1000~\rm km~s^{-1}$ of the redshift of the QSO CTS G18.01. Four of these X-ray sources are associated with CO(4-3) emitting-line galaxies (i.e., ID1, ID2, ID3, ID6), and ID3 is also associated to a CO(9-8) emission line \citep{Pensabene24}. 

We also expanded our research on X-ray emission in protocluster members by including candidates selected with FORS as LBGs at $3<z<3.5$ using the UBR-selection method (\citepgalbiati). These sources are located within the FORS FoV shown as a magenta square in Figure~\ref{fig:ChandraField}.
We found that only one source out of 178, representing 0.6$\%$ of the sample, is associated with an X-ray detection with an offset of less than 0.5 arcseconds.
This single case is located outside the MUSE FoV, positioned 3.6 arcmin away from CTS G18.01. The remarkably low percentage of X-ray counterparts associated with photometrically-selected LBGs at $z \approx 3-3.5$ is due to the ineffectiveness of the UBR-selection method to include LBGs that host an AGN (\citepgalbiati). Indeed, none of the spectroscopically-selected galaxies associated to an X-ray emission is also selected as an LBG.
Three X-ray sources are identified within the MUSE FOV and are not associated to any spectroscopically-selected protocluster member (see left panel in Figure~\ref{fig:Detections}). One of these sources is linked to a galaxy with clear emission lines, constraining its redshift to $z \simeq 3.8$. The other sources are linked to galaxies that do not exhibit high-S/N emission lines.

In summary our cross-matching analysis revealed X-ray emission in six spectroscopically-selected protocluster members, inside the MUSE FoV region of 4 arcmin$^2$ (equivalent to $\rm \sim 4~cMpc \times 4~cMpc$ at the redshift of the protocluster) and within $\pm 1000~\rm km~s^{-1}$ from the central QSO, including X-ray emission of the QSO itself.

\section{Spectroscopic Analysis of X-ray Sources in the MQN01 Protocluster} \label{sec:spectraanalysis}

\setlength{\tabcolsep}{9pt}
\begin{table*}[!t] 
\caption{Summary of properties of the X-ray source candidates and the results from the best spectral-fit models: (1) Identification of X-ray point sources associated with spectroscopically confirmed proto-cluster members, with the name of the CO line-emitting galaxy identified in \cite{Pensabene24} given in brackets; (2) Coordinates of the X-ray sources; (3) Redshifts determined with MUSE and/or ALMA data analyses; (4) Number of net counts; (5) Column Density from the best-fit; (6) Photon Index of the best-fit power-law; and (7) Luminosity in the 2-10 keV energy band.}
\label{table:AGNX}
\begin{tabular}{ccccccccc}
\hline
(1) & \multicolumn{2}{c}{(2)} & \multicolumn{2}{c}{(3)} & (4) & (5) & (6) & (7)   \\[3pt]
\hline
\hline
Name & R.A. & Dec & \multicolumn{2}{c}{redshifts} & Cnts & $N_H ^{pl}$ & $\rm \Gamma_{ph}$   & $\rm \log(\frac{L_{2-10 ~keV}}{erg~s^{-1}})$ \\
     & [deg] & [deg] & {\small MUSE} & {\small ALMA} & (0.5-10 keV)  & $\rm 10^{22} ~[cm^{-2}]$ & &   \\[3pt]
\hline    
ID1 (QSO) & 10.3809 & -49.6032 & 3.23647 & 3.2502  & 2778 &  $<1.65$ & $2.23_{-0.06}^{+0.06}$ & $45.79_{-0.02} ^{+0.02}$ \\[3pt]
ID2 (L02) & 10.3816 & -49.6161 & 3.24541 & 3.25081 & 453 &  $28.2_{-16.5}^{+20.0}$  & $1.9_{-0.4}^{+0.4}$ & $45.08_{-0.22}^{+0.27}$ \\[3pt]
ID3 (L04) & 10.3833 & -49.6052 & 3.24302 & 3.2456  & 29 &  $131_{-52}^{+64}$ & 1.8 & $44.26_{-0.19}^{+0.18}$ \\[3pt]
ID4 & 10.384  & -49.6057 & 3.24313 & --      & 19 &  $<25$ & 1.8 & $43.69_{-0.17}^{+0.19}$ \\[3pt]
ID5 & 10.3957 & -49.6116 & 3.24105 & --      & 47 &  $<33$ & 1.8 & $43.93_{-0.19}^{+0.19}$ \\[3pt]
ID6 (L01) & 10.3966 & -49.6199 & 3.24587 & 3.2451  & 17 &  $<32$ & 1.8 & $43.52_{-0.24}^{+0.20}$ \\[3pt] 
\hline
\hline
\end{tabular} 
\end{table*}

We performed a spectral analysis of the X-ray sources associated with MQN01 protocluster members.
At first, we extracted the spectra from circular regions with 2 times a radius defined as $r_{ext} = (2.55 \times 0.67791 - 0.0405083 ~\theta + 0.0535066 ~\theta^2) \rm ~arcsec$ in \citep{Rosati02} and references therein. There is an exception for the extraction of the spectrum of the blended sources ID3 and ID4, for which we used a radius $2r_{ext}/3$, that is smaller than half the distance between the two X-ray sources, to avoid contamination between the two sources. 
The background regions, free of sources, were manually selected near each respective source in a more extended circle.
Through the \texttt{specextract} tool, we obtained the ancillary response file (ARF) and the redistribution matrix file (RMF) at the source position, and the area- and exposure-weighted combination of source (.pi) and background (.bkg) spectra.
For the spectral fitting analysis we used \texttt{XSPEC}, by setting a C-statistic \citep{Arnaud96}. We adopted a simple model \texttt{tbabs*(zphabs*pow)} consisting on a power-law convolved with a Galactic and intrinsic absorption. The Galactic absorption \texttt{tbabs} is set to $N_{H,Gal} = 1.18 \times 10^{20}~\rm cm^{-2}$ according with the NASA HEASARC tool at the aim point coordinate within 0.1 deg. 
For the sources with net counts exceeding 100 (i.e., ID1 and ID2) we initially included an iron K-$\alpha$ line at the rest frame energy of 6.4 keV, left free to vary within the redshift uncertainties, and we set the slope $\Gamma$ of the intrinsic power-law free. In all cases, we observed that the iron K-$\alpha$ line is poorly constrained and, therefore, it was removed from the spectral model.
For the rest of the sources, we assumed an intrinsic power-law slope $\Gamma = 1.8$ and only the power-law normalization and hydrogen column densities ($N_H$) are free parameters.

Figure~\ref{fig:xspectra} in Appendix~\ref{app:zoomin} illustrates the spectra and best-fit models of these six X-ray sources. Table~\ref{table:AGNX} shows the coordinates and redshifts of the X-ray protocluster members estimated from MUSE and/or ALMA data, along with the best-fit properties derived from the spectral analysis. These properties include the intrinsic $N_H$, $\Gamma$, and the unabsorbed luminosities at the rest frame energy of 2-10 keV ($L_{2-10 \rm~keV}$), along with their respective 90$\%$ confidence levels. The luminosities were derived by applying a convolution model to the intrinsic power law by using the \texttt{clumin} tool in \texttt{XSPEC}. We observe that all X-ray protocluster members exhibit $\rm \log(L_{2-10 ~keV}/erg~s^{-1}) > 43.5$ which indicates their classification as AGN \citep[see][]{Ueda14}.

In summary, we found that all detected X-ray sources within the MQN01 protocluster are associated with AGN activity. Two of these sources exhibit an obscuration levels surpassing $N_H > 10^{23}~\rm cm^{-2}$, while for the remainder the obscuration is presumably lower, as indicated by unconstrained upper limit values of $N_H$. 
A more detailed X-ray spectral analysis of the central QSO, also including a thermal component, will be presented in the next paper. In this work, our spectral analysis is only finalized in obtaining unabsorbed luminosities for all the X-ray AGN.

\section{Spatial Distribution, Luminosity Function and Overdensity of X-ray AGNs in MQN01 and Other Protoclusters}\label{sec:AGNcompover}

\subsection{Comparative Analyses: AGN population in MQN01, SSA22 and Spiderweb protoclusters}\label{sec:protolucters}

Protoclusters have been identified across epochs spanning from $z \sim 1.6$ to $z = 8$ \citep[see][for a review]{Overzier16}. Studying their AGN population, including characterizing their distribution, spatial density and individual properties, offers insights into the evolution of SMBH growth in relation to the development of these large-scale overdense structures. It also is crucial to reveal the role played by potentially enhanced feeding processes in these regions in promoting SMBH accretion. 
These compelling questions motivated our exploration, in this first paper, of the collective properties of the X-ray AGN population in the $z \sim 3.247$ protocluster MQN01, including its spatial and velocity distribution, overdensity, AGN fraction, and SMBH accretion rates.
To investigate any connection between the environment and the AGN population, these characteristics need to be compared to those in the field and other protoclusters. This comparison is challenging due to the limited number of homogeneous and robust studies of AGN populations in protoclusters. For this purpose, we considered two well-studied protoclusters: SSA22 and Spiderweb.

The SSA22 protocluster \citep{Cowie94,Steidel98} provides an excellent comparative case with respect to MQN01, because of a similar redshift ($z \sim 3.09$). Furthermore, a 6 $\rm~arcmin^2$ area is covered with the VLT/MUSE instrument, and, similar to MQN01, these observations have revealed the presence of extended \lya\ filamentary structures \citep{Umehata19}. A significant difference is the absence of a brightest QSO in SSA22, which could serve as a proxy for the minimum potential well position of the overdensity.
The other protocluster that we included in this comparative analysis is Spiderweb. This is a structure centered on the bright high-redshift radio galaxy PKS 1138 at $z \simeq 2.156$ and represents a prototypical protocluster. Similar to the MQN01 and SSA22 fields, the Spiderweb protocluster also hosts an extended ($\gtrsim \rm 200~kpc$) \lya\ nebula \citep{Pentericci97}.
Both the SSA22 and Spiderweb fields have been extensively studied with multi-wavelength observations, revealing several members down to Mpc scales.
For instance, \cite{Steidel98} discovered a highly significant concentration of 15 LBGs in the SSA22 protocluster. A LAEs overdensity of about six was identified within an area of $\rm \simeq 8.8 \times 8.9 ~arcmin^2$ i.e., $\simeq 284~\mathrm{cMpc}^2$ \citep{Steidel00,Yamada12}. 
In the Spiderweb protocluster, an overdensity of spectroscopically confirmed LAEs \citep{Kurk00,Pentericci00} and H$\alpha$ emitters \citep[HAEs;][]{Hatch11,Kuiper11,Shimakawa15,Perez23} has been found, covering areas of $\sim 111~\mathrm{cMpc}^2$.
Conversely, in the MQN01 field, an overdensity of spectroscopically confirmed protocluster members has been identified within the MUSE-mosaic FOV, covering approximately $4\times 4~\rm cMpc^2$ (equivalent to $\rm \sim 2 \times 2~arcmin^2$), indicated by multiple high-S/N absorption and emission features (\citepgalbiati; see Section~\ref{sec:multiW} for details).
Therefore, the differences in the selection methods for identifying members and the different spectroscopic data coverages in these three protoclusters prevent the possibility to use consistent physical method to derive comparable volumes, and imply that the comparison should be taken with some caution. 

On the other hand, deep Chandra observations available for these protoclusters, specifically with a total exposure time of approximately $\sim \rm 400~ks$ for SSA22 \citep{Lehmer09} and $\sim \rm 700~ks$ for Spiderweb \citep{Tozzi22}, provide similar hard flux sensitivity to those in this paper, as shown in  Figure~\ref{fig:SK}. \cite{Lehmer09} identified ten AGNs in SSA22 associated with LAEs and LBGs. Six of these AGNs are associated with spectroscopically selected protocluster members at $z \sim 3.1$, while two are located well beyond $\rm \pm 15000 ~km~s^{-1}$ from the protocluster redshift, and the remaining two are associated with only photometrically selected LBGs.
\cite{Tozzi22} identified 14 X-ray AGNs in Spiderweb, including the central radio-jet galaxy PKS 1138 \citep{Roettgering94}. One of these AGNs is a candidate member based on X-ray data without spectroscopic confirmation, and two are Compton-Thick candidates, for which only the absorbed luminosity can be provided.
Table~\ref{table:PProt} reports various properties of these protoclusters, of their AGN populations and the exposure times of the respective Chandra observations. In the 4th column, we report the number of X-ray AGNs, associated to a spectroscopically-selected galaxy within $\pm 4000~\rm km~s^{-1}$ from the protocluster redshift (3rd column), and in the brackets we report those associated to photometrically-selected protocluster member candidates.

\begin{table}[t] 
\caption{Physical properties of the MQN01, SSA22, and Spiderweb protoclusters and their AGN populations: (1) Name; (2) Chandra Exposure Time; (3) Redshift of the Protocluster; (4) Number of spectroscopically selected X-ray AGNs ($\rm N_{AGN;4000}^{spec-z}$) within $\pm 4000 \rm~km~s^{-1}$ of the protocluster redshift, along with the number of photometrically selected AGNs ($\rm N_{AGN}^{phot-z}$) shown in brackets; (5,6) Minimum velocity range ($\rm  \Delta v_{AGN}^{spec-z}$) and circular area ($A_{AGN;4000}^{spec-z}$) encompassing all spectroscopically selected AGNs within the protocluster; (7) Maximum number of AGNs within a area of $\sim 16~\rm cMpc^2$ and within $\pm 1000~\rm km~s^{-1}$ from the protocluster redshift.}
\begin{adjustbox}{width=1\columnwidth,center}
\label{table:PProt}
\begin{tabular}{@{}c@{$~~$}|c@{$~~$}c@{$~~$}c@{$~~$}c@{$~~$}c@{$~$}c@{}}
\hline
(1) & (2) & (3) & (4) & (5) & (6) & (7)  \\[3pt]
\hline
\hline \\[-9pt]
Name & $t_{\rm exp}$ & $z_{prot}$ & $\rm N_{AGN;4000}^{spec-z}$ & $\rm  \Delta v_{AGN}^{spec-z}$ & $A_{AGN;4000}^{spec-z}$  & $N_{AGN}^{V_{MQN01}}$  \\[5pt]
    & ks &  & ($\rm N_{AGN}^{phot-z}$) & $\rm [km~s^{-1}]$ & $\rm [cMpc^2]$ &    \\[2pt]
\hline \\[-9pt]    
MQN01 & 634      & 3.247 & 6 (0)   & 689 & 4.2 & 6  \\
SSA22 & 400      & 3.092 & 6 (2) &  1610 & 463 & 2   \\
Spiderweb & 700  & 2.156 & 13 (1) & 1830 & 77 & 7 \\[3pt]
\hline
\hline 
\end{tabular}
\end{adjustbox}
\end{table}

\subsection{AGN Spatial Distribution within Protoclusters}\label{sec:AGNcompact}

The compactness of the AGN population in different environments provides insight into SMBH feeding processes and the potential extent of their feedback effects. 
We wondered whether a same compactness of AGN found in MQN01 was also found in the other two protoclusters. To address this question, our assumption was that the center of these protoclusters coincides with the region of highest galaxy overdensity.

\begin{figure}[t]
   \begin{center}   
   \includegraphics[height=0.4\textheight,angle=0]{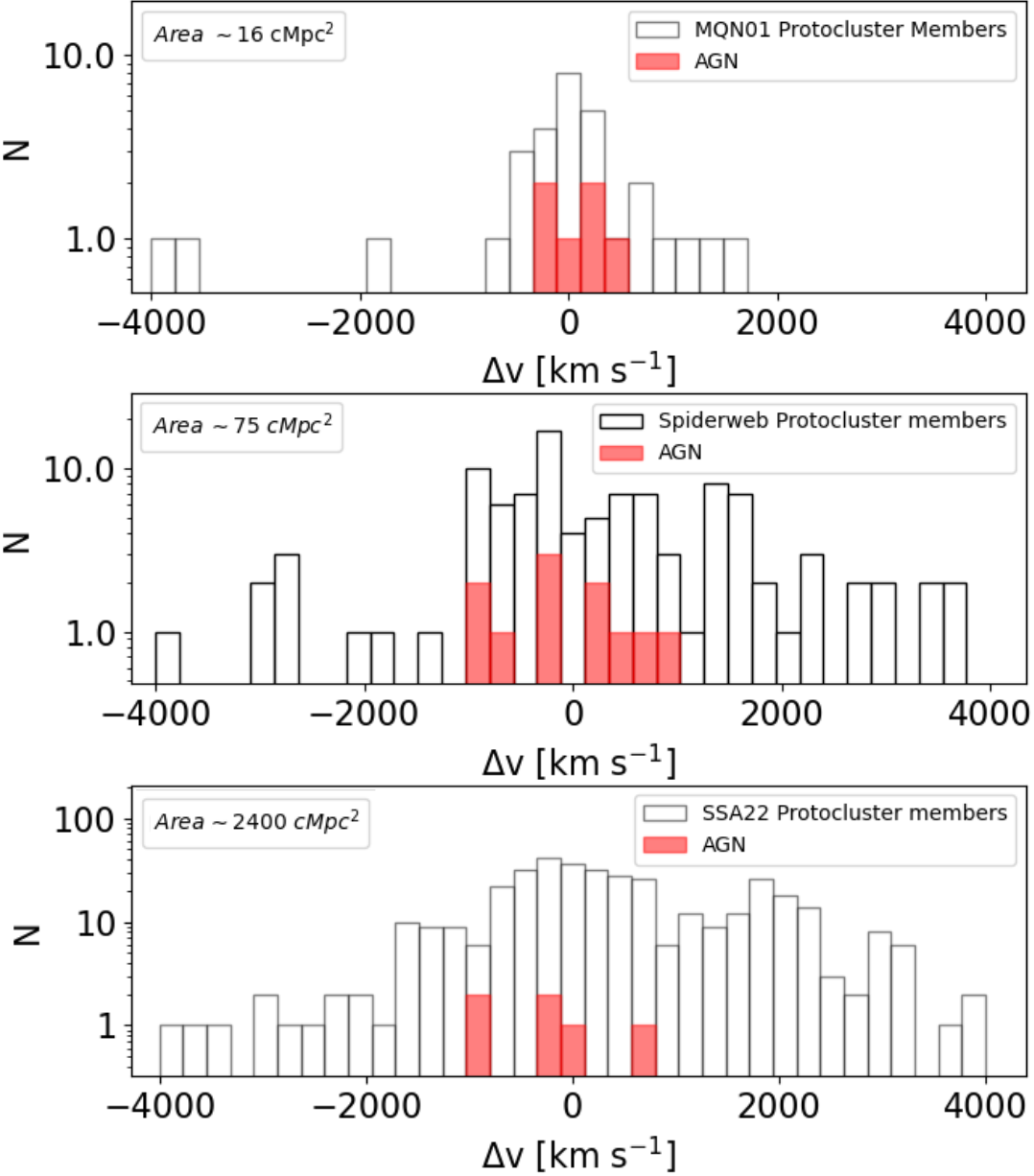}
   \caption{Velocity distribution of the galaxies+AGN (white) and AGN (red) members of the MQN01 (top panel), Spiderweb (middle panel), and SSA22 (bottom panel) protoclusters, in a velocity range $\pm 4000\rm ~km~s^{-1}$ and within areas of approximately 16, 75 and 2400$\rm ~cMpc^2$, respectively, as reported in the top left legend of each panel.}
   \label{fig:zdist}
   \end{center}
\end{figure}

\begin{figure}[t]
   \begin{center}   
   \includegraphics[height=0.77\textheight,angle=0]{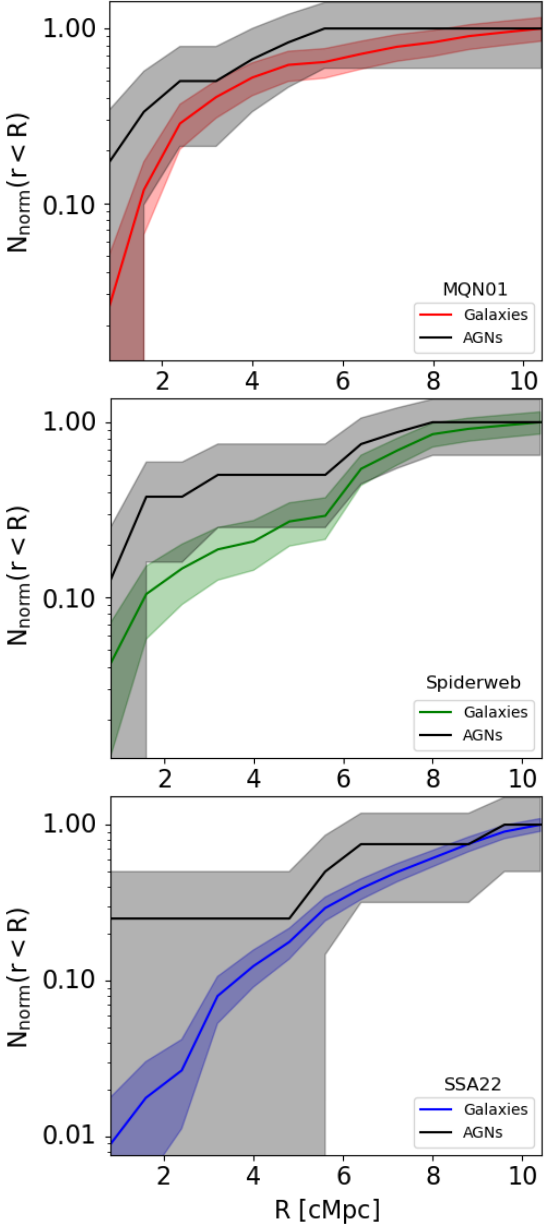}
   \caption{Radial profiles, normalized to their maximum values, of the cumulative number of galaxies and AGNs within a sphere of radius $\rm r<R$ for different protoclusters. The top, middle and bottom panels display these profiles for galaxies (red, green, blue) and AGNs (black) in MQN01, Spiderweb, and SS22, respectively. The transparent bands represent the Poissonian uncertainties in the number of objects within each sphere.}
   \label{fig:Ncum}
   \end{center}
\end{figure}

The histograms in Figure~\ref{fig:zdist} illustrate the velocity distribution of spectroscopically-selected galaxies (white) and those hosting AGNs (red) within $\pm 4000\rm ~km~s^{-1}$ of the redshift of each protocluster. These distributions are shown for galaxies within the specified areas indicated at the top left of each panel for each protocluster. Naturally, a larger survey area from which protocluster member galaxies are selected leads to a greater number of galaxies and a broader velocity distribution.
The top panel show the case for MQN01, the middle panel displays data for the Spiderweb protocluster, with galaxies detected in several studies \citep{Pentericci00,Kurk04b,Croft05,Kuiper11,Tanaka13,Dannerbauer14,Shimakawa18,Jin21}, and the bottom panel shows the distribution of galaxies associated with the SSA22 protocluster as collected by \cite{Mawatari23}.
The redshifts of the protoclusters were obtained as the mean redshift of all galaxies within a $\pm 1000 \rm~ km~s^{-1}$ range from an initial protocluster redshift, i.e., 3.25, 3.09, and 2.156 for MQN01, Spiderweb, and SSA22, respectively, according to the literature. These estimates allow an excellent centering on the peak of galaxies in velocity distribution and are listed in the 3rd column in Table~\ref{table:PProt}. 
The AGN velocity distribution in MQN01 appears being more compact compared to those in SSA22 and Spiderweb, with  velocity ranges of $ 689 ~ \text{km~ s}^{-1}$, $1610 ~ \text{km~ s}^{-1}$, and $1830 ~ \text{km ~s}^{-1}$, respectively (as reported in the 5th column of Table~\ref{table:PProt}).
After that, we derived the spatial distribution of the AGNs within $\pm 4000~\rm km~s^{-1}$ from the protocluster redshift by measuring the area of circles with diameters equal to the maximum projected distance between X-ray AGNs. We found areas of approximately 4.2, 463, and 77$\rm ~cMpc^2$ for the MQN01, SSA22, and Spiderweb protoclusters, respectively (see the 8th column of Table~\ref{table:PProt}).
These estimates of velocity and spatial distributions reflect the maximum extent of the AGN populations found in the three protoclusters, which is largely dependent on the available observations. 

To evaluate the compactness of active galaxy populations independently of the spectroscopic survey areas and the overall distribution of the AGN population, we counted the number of AGNs within a specific volume across all protoclusters. 
We defined this volume as a cylinder with a cross-sectional area of $\simeq 16~\text{cMpc}^2$ and a velocity range of $\pm 1000 ~\text{km~s}^{-1}$, representing the space that contains the majority of galaxies associated with the MQN01 protocluster. 
This approach ensures that the volume is independent of AGN distribution. 
Additionally, to avoid bias introduced by the method used to define the protocluster’s center, we positioned the volume to maximize the number of AGNs it contained within each protocluster. 
By construction, this volume encompasses six AGNs in the MQN01 protocluster. 
In comparison, we found a maximum of seven AGNs in the Spiderweb protocluster and two AGNs in the SSA22 protocluster (see the 7th column of Table~\ref{table:PProt}). 
According to \cite{Gehrels86}, the Poissonian intervals corresponding to a 1$\sigma$ Gaussian confidence level for these numbers of AGNs are [3.62, 9.58] for MQN01, [4.42, 10.8] for Spiderweb, and [0.71, 4.64] for SSA22. Thus, these values are consistent within the Poissonian errors.
Specifically, the Poissonian probability of observing exactly six (seven) AGNs, as in MQN01 (Spiderweb), when two (six) AGNs are expected, as in SSA22 (MQN01), is about 1.2$\%$ (14$\%$). For comparison, if seven (six) AGNs are detected, the Poisson probability that this is the actual number of AGNs is 15$\%$ (16$\%$).\\

What role does galaxy compactness play in shaping the compactness of AGNs? The observed compactness of AGNs could be a result of a denser galaxy population in certain regions, or it might reflect an intrinsic characteristic of the distribution of nuclear activity. In an attempt to distinguish between these possibilities, in Figure~\ref{fig:Ncum}, we compare the radial profiles of the cumulative number of galaxies (red, green, and blue lines) and AGNs (black lines) within a sphere of radius $\rm r<R$, normalized by the total number of objects (hereafter $N_{norm}(r<R)$). These profiles are shown for MQN01 (top panel), Spiderweb (middle panel), and SSA22 (bottom panel), using reference coordinates and redshifts that maximize the values in the first radial bins of the $N_{norm}(r<R)$ profile, following the approach described earlier. Our finding show that the average $N_{norm}(r<R)$ of AGNs (indicated by black solid lines) is systematically higher towards the center compared to that of galaxies, although these are consistent within the Poissonian uncertainties (represented by the shaded areas).
We applied the Anderson-Darling test to pairs of $N_{norm}(r<R)$ profiles for AGNs and galaxies within each protocluster to assess whether, within the uncertainties, they could originate from the same distribution. 
We utilized the Python tool \texttt{anderson\_ksamp} \citep{Scholz87}, incorporating a bootstrap algorithm to evaluate all profiles within the Poissonian uncertainties.
The results were inconclusive: in over 89$\%$ of the cases, we could not reject the hypothesis that the two samples originate from the same distribution, suggesting that they are not statistically distinct. Consequently, due to the limited sample size and the associated Poissonian uncertainties, drawing definitive conclusions about the relative compactness of AGNs and galaxies remains challenging.


\subsection{AGN Overdensity in Protoclusters: An Excess of the Brightest AGNs in MQN01}\label{sec:XLF}

In this section, we aimed at quantifying the AGN space density and overdensities across different luminosity ranges in MQN01 protocluster, and compare these with the same estimates in SSA22 and Spiderweb protoclusters presented in Section~\ref{sec:protolucters}. To do it, we derived the X-ray luminosity function (XLF), that is a optimal tool to study the population of actively accreting SMBHs and to evaluate the density of such occurrences in specific parts of the Universe.
To date, there is limited research on the study of AGN in high-redshift protoclusters, and more specifically, on the estimation of the XLF and the comparison with that observed in the field at the same epoch. 
\cite{Krishnan17} first showed the XLF at 0.5-7 keV for the protocluster Cl 0218.3-0510 at z=1.62 compared to that in the field. Their analysis led to the conclusion that the XLF is systematically higher in the protocluster with no clear evidence indicating discrepancies in the accretion rate distributions between AGN in the field and those within the protoclusters. Other cases are reported in the recent papers such as \cite{Tozzi22}, in which the cumulative XLF in the Spiderweb protocluster and in the field are shown. This comparison shows an increasing of the overdensity of AGN with 2-10 keV X-ray luminosities exceeding $\rm \sim 10^{44}~erg~s^{-1}$ in Spiderweb, at each energy band. Even in \cite{Vito24}, the space density of AGN derived from the combination of two identical extreme protoclusters at $z \sim 4.3$ is shown. In this study, the detection of two X-ray AGN whose $L_{2-10~\rm keV}$ exceed $\rm 10^{45}~erg~s^{-1}$, imply an increase of a three/five order of magnitude in the XLF at the luminosity bin $\log(\rm L_{2-10~keV}/erg~s^{-1}) = 45-46$ compared to the field at the same redshift.\\

We provide a novel comparison of the hard (2-10 keV) XLF (hereafter HXLF) of AGN estimated for a small sample of protoclusters, i.e. MQN01, studied in this work, SSA22 and Spiderweb, and for the fields at the same redshifts. 
To obtain the cumulative HXLF in a grid of 2-10 keV X-ray luminosities (hereafter $L_X$), we estimated the space density ($\phi$; in $\rm Mpc^{-3}$ units) of AGNs with luminosity exceeding $L_X$, using the $\rm 1/V_{max}$ method \citep{Schmidt68,Avni80,Vito14}:
\begin{equation}\label{eq:Phi}
	\phi (L>L_X) = \sum _{i=1} ^{N(L_i>L_X)} \frac{1}{\int _{z_{min}} ^{z_{max}} (\Omega(L_i,z)/4\pi) (dV/dz) dz } 
\end{equation}
where $N(L_i>L_X)$ is the number of AGNs with luminosity above $L_X$ and with spectroscopic redshift within $\Delta z$, and $\Omega(L_i,z) \equiv \Omega(flux)$, with units in steradians, represents the sky coverage shown in Figure~\ref{fig:SK} integrated over redshift and luminosity. The range $\Delta z$ corresponds to the velocity range $\pm \Delta v_{max}$, around the redshift of the protocluster, defining the volume within which we estimated the overdensity of AGN.
The term $\Omega(L_i,z)/4\pi$ accounts for the fraction of sky covered by X-ray AGN above a given $L_X$, and represents the area sensitive to Chandra observations above a flux threshold. We set, as maximum value of $\Omega$ a specific area (hereafter $A_{max}$) associated with the volume in which we want to estimate the overdensity of AGN in the protocluster, and this area may be particularly important for the HXLF brightest end.  
Breaking down the denominator of equation (\ref{eq:Phi}), we obtain the following equation:
\begin{equation}
	(dV/dz)dz = \sum _{j=1} ^{n(z_{bins})-1} \frac{\Omega(L,z_j)}{3} [D_{comov}^3 (z_{j+1}) - D_{comov} ^3 (z_j)]
\end{equation}
where $\Omega(L,z_j)$ is multiplied by the differences between the cubic comoving distances in different redshift bins. This represents the volume of a spherical shell between $z$ and $z+\Delta z$ in a specific fraction of the solid angle $\Omega(L,z_j)/4 \pi$. 
To be conservative we did not apply the incompleteness correction as done in \cite{Vito18}.
To estimate the uncertainties of $\phi$, we performed iterations of this calculation by randomly varying both the number of detected X-ray AGNs according to the Poisson statistics, that dominate the error, and the unabsorbed luminosities, assuming a Gaussian distribution centered around the mean luminosity up to extreme values marked by the $90 \%$ confidence level values reported in Table~\ref{table:AGNX}.\\

\begin{figure}[t]
   \begin{center}   
   \includegraphics[height=0.31\textheight,angle=0]{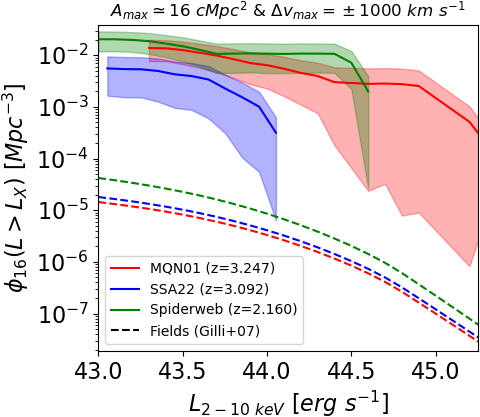}
   \caption{Cumulative HXLF of AGNs in a volume ($V_{16}$) defined to contain the bulk of the galaxy population in MQN01, with $A_{max} \sim \rm 16~cMpc^2$ and $\Delta v_{max} = \pm \rm 1000~km~s^{-1}$. The red and green lines represent the HXLF estimated for the MQN01 and Spiderweb protoclusters, respectively, centered on the brightest QSOs, CTS G18.01 (ID1) and PSK1138, which are excluded from the analysis. The blue line depicts the HXLF of AGNs in the SSA22 protocluster, centered at the location that maximizes the AGN overdensity within $V_{16}$. The transparent bands indicate the range of values due to Poisson uncertainties. The dashed lines indicate the HXLF of AGNs in the field, estimated as in \cite{Gilli07}, with colors corresponding to the same redshifts as the solid lines for the protoclusters.}
   \label{fig:xlfAMQN} 
   \end{center}
\end{figure}

\begin{figure}[t]
   \begin{center}   
   \includegraphics[height=0.31\textheight,angle=0]{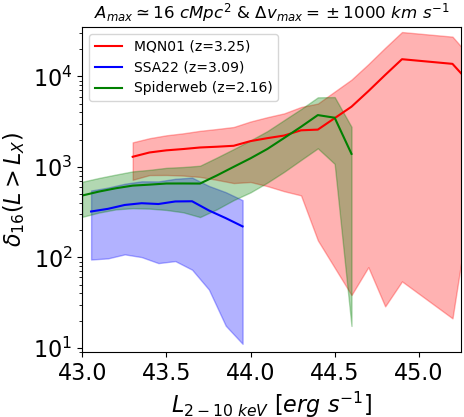}
   \caption{Cumulative luminosity function of the overdensity of X-ray AGN derived as the ratio between the $\phi_{16}$ shown in Figure~\ref{fig:xlfAMQN} and the $\phi$ derived for the Field according to \cite{Gilli07}, normalized by the latter. The red, blue, and green curves represent the overdensities for MQN01, SSA22, and Spiderweb, respectively.}
   \label{fig:OverDen}
   \end{center}
\end{figure}

The values of $\phi$ depend on the volume defined by $A_{max}$ and $\Delta v_{max}$, as discussed above. The selection of this volume for each protocluster is critical for making appropriate comparisons of AGN densities within a given galaxy overdensity. 
However, as discussed in Section~\ref{sec:protolucters}, these protoclusters lack uniformly extensive spectroscopic coverage and employ varying methods for selecting galaxy populations. This disparity impacts both the maximum areas surveyed for AGN identification and the respective completeness, thereby affecting the perceived overdensity of the AGN populations.
Thus, our primary objective was to determine whether the values of $\phi$ and the AGN overdensities observed in MQN01, across different $L_X$, are also present in the Spiderweb and SSA22 protoclusters, regardless their geometrical and volumetric differences.

To achieve this, we defined the maximum volume as the one previously used in Section~\ref{sec:AGNcompact}, encompassing the majority of galaxies in MQN01. Specifically, this volume has a cross-sectional area of $A_{max} = 16~\rm cMpc^2$, corresponding to the MUSE FOV, and a velocity range of $\Delta v_{max} = 1000~\rm km~s^{-1}$ (see Figure~\ref{fig:zdist}). This volume will be referred to as $V_{16}$ hereafter.
To ensure consistency with usual methods used in literature \citep[e.g.,][]{Tozzi22} to estimate the HXLF, $V_{16}$ was centered on the positions of the most luminous QSOs in the MQN01 and Spiderweb protoclusters, specifically CTS G18.01 (ID1) and PSK1138, respectively. These QSOs were excluded from the calculation.
For the SSA22 protocluster, given the lack of a brightest QSO, $V_{16}$ was centered at the location that maximizes the AGN density, following the approach used in Section~\ref{sec:AGNcompact}. 
Furthermore, to derive the HXLF in Spiderweb we assumed that the absorbed luminosity in the two Compton-Thick cases is $\sim$10$\%$ of their intrinsic luminosity, as done in \cite{Tozzi22}. For SSA22, we used the 2-8 keV X-ray luminosities reported in \cite{Lehmer09} and multiplied by a factor of approximately 1.18 to convert these to 2-10 keV X-ray luminosities. This conversion is based on the assumption that the bulk of the AGN models are consistent with a $\Gamma=1.8$ power-law.\\

Figure~\ref{fig:xlfAMQN} depicts the $\phi_{16}(L>L_X)$ derived, in $V_{16}$, in the MQN01 (red line), SSA22 (blue) and Spiderweb (green) protoclusters across a range of $L_X$ of the AGNs. The shaded regions represent Poissonian uncertainties, dominated mainly by Poissonian errors on the number of AGN. Dashed lines of different colors indicate the AGN cumulative HXLF in the field, as estimated by \cite{Gilli07}, at the same redshifts as the solid lines for the protoclusters. These HXLF are consistent with those derived in other studies \citep[e.g.,][]{Ueda14,Vito14,Georgakakis15,Ranalli16,Vito18,Wolf21}

To estimate the overdensity of AGN within $V_{16}$ (hereafter $\delta_16$) in protoclusters at different redshifts, as a function of the X-ray luminosity, we adopted the the following formula:
\begin{equation} \label{eq:delta}
	\delta _{16} = \frac{\phi(L>L_X)_{16} ^{Prot} -\phi(L>L_X) ^{Field}}{\phi(L>L_X) ^{Field}}
\end{equation}
where $\phi(L>L_X)_{16} ^{Prot}$ is the AGN space densities obtained from equation (\ref{eq:Phi}) and $\phi(L>L_X) ^{Field}$ is that shown in Figure~\ref{fig:xlfAMQN} as derived in \cite{Gilli07}.
Figure~\ref{fig:OverDen} illustrates these $\delta_{16}$ estimates for the AGN in the MQN01 (red), SSA22 (blue), and Spiderweb (green) protoclusters.\\

The HXLF of the AGN population in protoclusters, at least within a volume $V_{16}$, is consistently higher than that observed in the field, aligning with findings in the literature \citep{Krishnan17,Tozzi22,Vito24} and supports the increased frequency of AGN activity in these overdense environments.
Additionally, the average curves for MQN01 and Spiderweb, indicated by solid lines, are flatter compared to those observed in the field at similar redshifts. This observation is clearer in Figure~\ref{fig:OverDen}, which shows a growth of $\delta_{16}$ as a function of $L_X$ for MQN01 and Spiderweb. 
Such a trend is stronger in Spiderweb up to $\log(\rm L_{2-10 ~ keV}/ erg~s^{-1}) \approx 44.4$, and continues in MQN01 up to $\log(\rm L_{2-10 ~ keV}/ erg~s^{-1}) \approx 45.3$, although with less significance due to Poissonian errors.  
Moreover, the overdensity of $\delta_{16} \sim 10^2-10^4$ at $\log(\rm L_X/ erg~s^{-1}) > 45$ obtained assuming a volume $V_{16}$, aligns with $\delta$ in the range of three to five orders of magnitude at the luminosity bin $\log(\rm L_{2-10keV} / \rm ergs^{-1}) = 45-46$ found for two $z \sim 4.3$ protoclusters in \cite{Vito24}. 
This significant overdensity in MQN01 underscores the rarity of discovering two extremely luminous AGNs within approximately $\sim \rm 1.5~cMpc$ of each other, making the type 2 AGN, ID2, particularly noteworthy.
This provides clear evidence of a significant increase in the brightness of the AGN XLF in protoclusters.

As will be discussed in Section~\ref{sec:AGNfraction}, the high AGN overdensity $\delta_{16}$ in the MQN01 protocluster is not directly related to the protocluster environment's effect on promoting SMBH growth. Indeed, it may be a consequence of an equally high density of galaxies, highlighting the importance of estimating an AGN fraction. However, when comparing the result $\delta_{16}(\text{AGN}) > 10^3$ with the galaxy overdensity $\delta_{16} (\text{Galaxy}) \simeq 53$ within the same volume\footnote{$\delta_{16} = 53 \pm 17.4$ at $\rm M_R \leq$ -19.25 mag}, as reported in the upcoming paper \citegalbiati, it becomes evident that the AGN overdensity cannot be solely attributed to a large density of galaxies.

\section{Protocluster Environments and Their Role in SMBH and Galaxy Evolution}\label{sec:envvsSMBH}

In this section, we study the relation between the properties of the detected X-ray AGN in MQN01 and the associated galaxy population in order to understand the possible effect of environment on AGN triggering and SMBH growth. In particular, we focused on the AGN fraction as a function of galaxy stellar mass ($M_{*}$) and on the SMBH accretion rate ($\lambda _\text{sBHAR}$). The value of these two parameters obtained for the MNQ01 field are compared to other protoclusters, such as SSA22 and Spiderweb and to the field.

\subsection{AGN Fraction in MQN01: Ubiquity of Accreting SMBHs in Massive Galaxies}\label{sec:AGNfraction}

\begin{table}[t] 
\caption{Properties of the AGN host-galaxies derived from the SED-fitting analysis in \citegalbiati.}
\begin{adjustbox}{width=1\columnwidth,center}
\label{table:HostProp}
\begin{tabular}{@{}c@{$~~~$}c@{$~~~$}c@{$~~~$}c@{$~~~$}c@{}}
\hline
\hline 
Name & $\log(\text{SFR}/ \rm M_{\odot}~yr^{-1})$ &  $\log(\text{sSFR}/ \rm yr^{-1})$ &  $\log[M_{*}/M_{\odot}]$ &   $A_V$  \\
\hline    \\[-7pt]
ID2 & $2.26 \pm 0.32$ & $-8.92 \pm 0.39$ & $11.19 \pm 0.22$ & $1.15 \pm 0.05$ \\[3pt]
ID3 & $2.19 \pm 0.33$ & $-8.89 \pm 0.40$ & $11.08 \pm 0.22$ & $0.88 \pm 0.06$ \\[3pt]
ID4 & $1.70 \pm 0.38$ & $-8.92 \pm 0.43$ & $10.62 \pm 0.22$ & $1.70 \pm 0.13$ \\[3pt]
ID5 & $1.24 \pm 0.38$ & $-8.75 \pm 0.44$ & $9.99 \pm 0.23$  & $0.90 \pm 0.06$ \\[3pt] 
ID6 & $2.09 \pm 0.30$ & $-9.28 \pm 0.36$ & $11.37 \pm 0.20$ & $0.58 \pm 0.03$ \\[3pt]
\hline
\hline \\
\end{tabular}
\end{adjustbox}
\end{table}

\begin{figure*}[t]
   \begin{center}
   \includegraphics[height=0.44\textheight,angle=0]{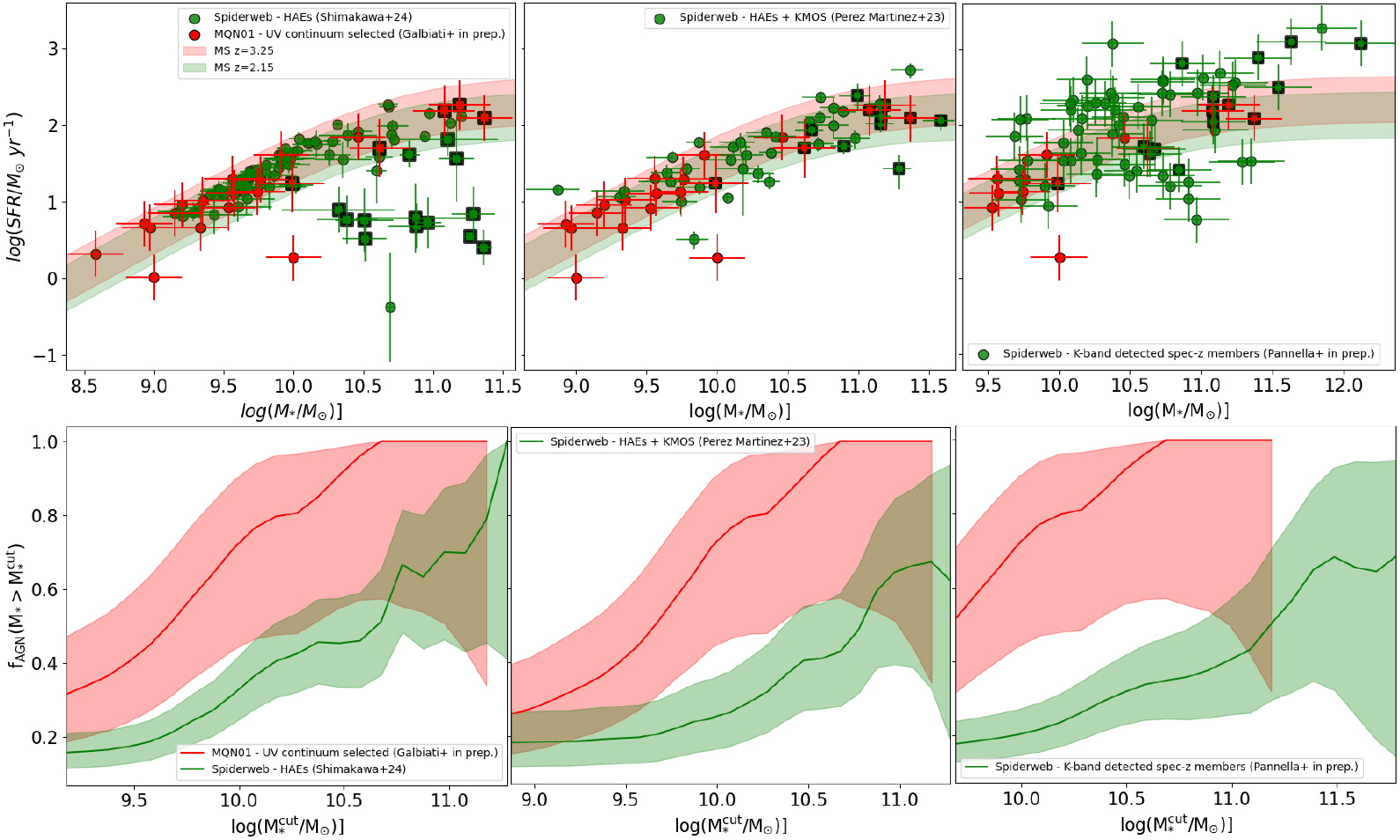}
   \caption{The top panels display the $\text{SFR}$ versus $M_*$ plots for galaxies in the MQN01 (red dots) and Spiderweb (green dots) regions. The Spiderweb data were obtained by \cite{Shimakawa24} (left panel), \cite{Perez23} (middle panel), and \citepannella\ (right panel). The shaded areas represent the \cite{Popesso23} main sequence (MS) with a $\pm 0.3~dex$ spread at redshifts $z=3.25$ (red) and $z=2.16$ (green). Galaxies hosting an X-ray AGN are highlighted with black-edged squares. The bottom panels show the variation of $f_{AGN}$ as a function of the $M_*$ threshold, as indicated by the grid of values on the x-axis.}
   \label{figure:MSFR}
   \end{center}
\end{figure*}

One of the most straightforward parameters that can help us understand the impact of the protocluster environment on promoting SMBH growth is the fraction of galaxies within a given population that host X-ray AGNs, denoted as $f_{AGN}$. However, comparing $f_{AGN}$ across different protocluster fields is challenging due to variations in galaxy survey selection techniques and their associated completeness. Both galaxy and AGN selections are influenced by parameters such as stellar mass ($M_{*}$), star formation rate (SFR), and AGN X-ray luminosity. Additionally, different methods for determining stellar masses and SFRs result in rather heterogeneous samples of galaxies as a function of $M_{*}$ and SFR across various fields.
Consequently, the $f_{AGN}$ values reported in the literature for different protoclusters \citep{Tozzi22,Vito23,Vito24} are not easily comparable. Nonetheless, these values, typically around or above $10^{-2}$, are significantly higher than those measured in the field \citep[$< 10^{-2}$][]{Martini13}. To illustrate how a single parameter can affect $f_{AGN}$, we defined $f_{AGN}$ as a function of the stellar mass cut ($M_{*}^{cut}$) of the galaxy sample used to estimate it.

In order to take into account these challenges, we limited our analysis to the MQN01 and Spiderweb fields for which multiple estimates of $M_{*}$ and $\text{SFR}$ are available. Specifically, for Spiderweb, we used values from \cite{Shimakawa24}, \cite{Perez23}, and an ongoing study by \citepannella, each of which considers different samples obtained with different methodologies. While $M_{*}$ is generally well constrained through SED fitting, the estimation of $\text{SFR}$ is more complex, given its dependence on recent star formation history, dust extinction, and the age of the stellar population. 
\cite{Shimakawa24} performed SED fitting using the \texttt{X-CIGALE} tool \citep{Yang20} with photometric data from X-ray to submillimeter for a sample of 84 HAEs based on narrow-band samples and BzKs color selection \citep{Daddi05,Shimakawa18}. 
For a sample of spectroscopically selected protocluster members, using KMOS data, \cite{Perez23} applied a similar SED-fitting procedure for $M_{*}$ derivation, while the $\text{SFR}$ was derived using the \cite{Kennicutt98} formula modified for a \cite{Chabrier03} IMF.
They showed that $\text{SFR}$ values from SED-fitting align with those from \ha\ emission lines.
Finally, \citepannella\ uses the multi-band catalog described in \cite{Tozzi22} to estimate stellar masses through SED fitting \citep[\texttt{fastpp} tool\footnote{https://github.com/cschreib/fastpp} in][]{Kriek09} and SFRs by correcting the observed FUV rest-frame emission according to eq.10 in \cite{Pannella15}, i.e. the correlation between UV dust attenuation and galaxy stellar mass.

Galaxies in MQN01 have been selected as discussed in Section~\ref{sec:MUSE} using the deep available MUSE data providing a complete census of galaxies with spectroscopic data above a given rest-frame UV magnitude of 26.5 (at the rest-frame reference wavelength of 150 nm). As such, they represent a population of not heavily dust-obscured star forming galaxies. We note however, that the deep ALMA observations of \cite{Pensabene24} did not reveal additional massive and dusty galaxies in the field with the exception of a source located at about 1'' from the QSO which is not present in either the MUSE or the Chandra sample due to the large QSO PSF in both datasets. The $M_*$ and $\text{SFR}$ have been obtained using a method similar to the one used in \cite{Shimakawa24} and are listed in Table~\ref{table:HostProp} for galaxies identified as hosting an X-ray AGN.

The top panels of Figure~\ref{figure:MSFR} present the $\text{SFR}$ versus $M_*$ distribution for galaxies in the Spiderweb protocluster (green) as obtained by \cite{Shimakawa24} (left panel), \cite{Perez23} (middle panel), and \citepannella\ (right panel), along with those in the MQN01 protocluster (red). The shaded areas indicate the \cite{Popesso23} main sequence (MS) with a $\pm 0.3~dex$ spread at redshifts $z=3.25$ (red) and $z=2.16$ (green). Galaxies hosting an X-ray AGN are marked with black-edged dots. In this analysis, we excluded the most luminous QSOs around which the protoclusters are centered, namely CTS G18.01 and the radio galaxy PSK1138, as they could represent outliers in the overall AGN and galaxy population. By comparing the positions of the Spiderweb galaxies across these three plots, the impact of galaxy selection and SFR estimation method on the evaluation of the galaxy population in the main sequence becomes evident. 

The bottom panels of Figure~\ref{figure:MSFR} display $f_{AGN}$ for all samples, calculated by including galaxies with $M_*$ above the thresholds indicated on the x-axis (i.e., $M_*^{cut}$). To estimate the 68$\%$ binomial error for $f_{AGN}$, we employed Jeffrey's Bayesian statistics using the Python tool \texttt{statsmodels}. The resulting uncertainties are represented by shaded areas \citep[for further details, see][]{Vito24}.

The evidence of a wide range of $f_{AGN}$ observed as a function of the $M_* ^{cut}$ emphasizes the need for consistent $f_{AGN}$ estimation across all protoclusters to enable direct comparisons. 
Despite these challenges some clear patterns are present. In all cases, $f_{AGN}$ increases with $M_* ^{cut}$, consistent with robust findings in the literature \citep[e.g.,][]{Kauffmann03,Best05,Haggard10}, which show that AGNs, with $L_X \gtrsim 10^{43.5}~ \rm erg ~s^{-1}$, are more frequently hosted by more massive galaxies. Recently, this trend has been extended to high-redshift ($2 < z < 4$) AGNs in dense environments identified through methods beyond just X-ray selection, as shown by \cite{Saha24}. Specifically, they demonstrate that the AGN fraction rises in higher-density regions and with increasing $M_*$. 
We note that the $f_{AGN}$ in MQN01 is larger than that observed in Spiderweb at all masses, regardless of the details of the $M_*$ determinations. Additionally, in the MQN01 protocluster, the average $f_{AGN}$ (solid line) reaches 1 for galaxies with $\log(M_*/\rm M_{\odot}) > 10.5$. In contrast, in Spiderweb, $f_{AGN}$ increases more gradually and reaches an average $f_{AGN}=1$ at $\log(M_/\rm M_{\odot}) > 11.5$ only using stellar masses as in \cite{Shimakawa24}, while $f_{AGN}$ in Spiderweb is always less than 1 with the other two methods. In other words, while Spiderweb contains massive galaxies without X-ray AGN counterparts, all galaxies in MQN01 exceeding $\log(M_*/\rm M_{\odot}) = 10.5$ and selected as discussed in section 3.1.1 are observed to host X-ray AGNs. 
On the other hand, the limited sample size in this study introduces substantial uncertainties, particularly in the estimates of $f_{AGN}$ at higher $M_{*}^{cut}$, which restricts our ability to fully interpret the impact of environments characterized by high gas and galaxy densities on the correlations between their properties and nuclear activity. 
However, assuming that both protoclusters have nearly complete selections of massive star-forming galaxies, the observations of a larger $f_{AGN}$ at all the masses, and that all galaxies above a certain mass host AGN in MQN01, but not in Spiderweb, could indicate an evolutionary trend in the corresponding growth of galaxies and their SMBHs. 
Specifically, this evidence suggests that the MQN01 protocluster at $z > 3$ may have more gas available around and within galaxies for BH growth (\citealt{Borisova16}; \citepcantalupo) and/or more efficient mechanisms for triggering SMBH accretion compared to the evolutionary stage and configuration of the Spiderweb protocluster. 
We also performed a similar analysis, focusing exclusively on star-forming galaxies, defined as those within $\pm 0.3~\rm dex$ of the Main Sequence (MS) as characterized by \cite{Popesso23}, thus excluding points outside the MS in the $\text{SFR}$ versus $M_*$ plots in Figure~\ref{figure:MSFR}.
This approach allows a comparative assessment of $f_{AGN} (M_* > M_*^{cut})$ between the MQN01 and Spiderweb star-forming galaxy populations. The results and conclusions from this analysis are largely unchanged.

\subsection{Impact of the Environment on the SMBH accretion rates}\label{sec:SMBHgrowth}

\begin{figure}[t]
   \begin{center}   
   \includegraphics[height=0.28\textheight,angle=0]{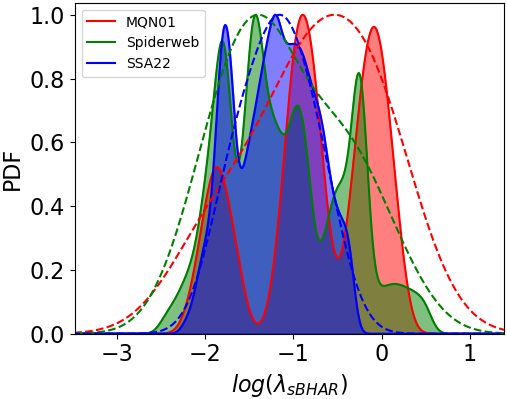}
   \caption{Probability Density Function (PDF) of the $\lambda_\text{sBHAR}$ parameter for the three SMBH populations host in the MQN01 (red), SSA22 (blue) and Spiderweb (green) protoclusters. Solid lines represent the $\lambda_\text{sBHAR}$ PDFs obtained using an iterative procedure that incorporates additional artificial sources, created by randomly varying $L_X$ and $M_*$ within their 1$\sigma$ uncertainties. Dashed lines show the $\lambda_\text{sBHAR}$ PDFs without applying this iterative procedure.}
   \label{fig:BHAR}
   \end{center}
\end{figure}

To provide a rough estimate of the growth rate of individual SMBHs in the populations of the MQN01, SSA22, and Spiderweb protoclusters, we estimated the parameter $\lambda _\text{sBHAR}$, following the methodology described in \cite{Bongiorno12} and \cite{Aird18}:
\begin{equation}\label{eq:BHAR}
	\lambda_{\text{sBHAR}} = \frac{k_{\text{bol}} L_X}{1.3 \times 10^{38} \times 0.002 \times M_*}
\end{equation}
where $L_X$ represents the 2-10 keV hard X-ray luminosity in erg/s, $k_\text{bol}$ is the bolometric correction factor to convert $L_X$ to the bolometric luminosity $L_{\text{bol}}$, and $M_*$ is the stellar mass of the galaxy in $\rm M_{\odot}$. Although $k_\text{bol}$ can vary as a function of both $L_\text{bol}$ and $L_X$ \citep{Marconi04,Bongiorno16,Duras20}, we adopted an average value $k_\text{bol}=25$, as used in \cite{Aird18}, because the uncertainties are primarily driven by other systematic errors, such as the assumption that $M_{BH}$ is equal to $0.002 \times M_*$ according to the relationship found by \cite{Haring04}.

We used the $L_X$ and $M_*$ values derived in the previous section for MQN01 and Spiderweb \citep[with the $M_*$ estimates taken from][]{Shimakawa24}, while the $M_*$ values for AGN host galaxies in SSA22 were taken from \cite{Monson23}. We derived $\lambda_\text{sBHAR}$ for the available sample of X-ray AGNs in each protocluster, excluding the luminous QSO CTS G18.01 in the MQN01 field and the radio galaxy PSK1138 in Spiderweb. 
To obtain the probability density function (PDF) of $\lambda_\text{sBHAR}$ for the SMBH populations in these different environments, we adopted the non-parametric Kernel Density Estimation (KDE) method \citep[see][and references therein]{Moon95}. 
The PDF of $\lambda_\text{sBHAR}$ for SMBH populations in MQN01 (red), SSA22 (blue) and Spiderweb (green) are depicted as dashed lines in Figure~\ref{fig:BHAR}. 

To enhance statistical robustness and better characterize the $\lambda_\text{sBHAR}$ distribution and properties of individual AGNs while accounting for measurement uncertainties, we employed an iterative approach using a bootstrap algorithm. This method involves expanding the initial sample by adding sources with $\lambda_\text{sBHAR}$ values derived from randomly varying the parameters $L_X$ and $M_*$ within their respective 1$\sigma$ uncertainties. This process effectively increases the sample size and provides a more comprehensive representation of the underlying distribution. The resulting $\lambda_\text{sBHAR}$ PDFs are shown in Figure~\ref{fig:BHAR} as solid lines subtended by colored areas.

The comparison of the peaks in the $\lambda_\text{sBHAR}$ PDFs represented by dashed lines, which can be considered as average estimates, suggests that MQN01 has a higher fraction of rapidly accreting SMBHs compared to the other protoclusters. 
How does this result compare to findings at different redshifts at in different environments?
Using deep X-ray observations from the XMM-LSS, COSMOS, and ELAIS-S1 extragalactic fields, \cite{Aird12} found that the probability for a galaxy to host an AGN with a given $\lambda_\text{sBHAR}$ ($p(\log(\lambda_\text{sBHAR})|M_*,z)$) follows a power-law distribution. This distribution is truncated at the Eddington limit, likely due to the self-regulation of SMBH growth \citep{Aird12,Bongiorno12,Aird18}. 
In a more recent study, \cite{Aird18} observed an evolution of the $p(\log(\lambda_\text{sBHAR})|M_*,z)$ for star-forming galaxies with $M_* > 10^{10}~\rm M_{\odot}$, noting that $\lambda_\text{sBHAR}$ shifts toward higher values at higher redshifts. This shift significantly impacts the increasing AGN duty cycle, defined as the fraction of galaxies with $\log(\lambda_\text{sBHAR})>0.1$. This trend could be explained by the greater access to a larger reservoir of cold gas for galaxies at higher redshifts. 
The average values of $\lambda_\text{sBHAR}$ found for galaxies in protoclusters i.e., the peaks of dashed lines in Figure~\ref{fig:BHAR}, align with the cosmic evolution of the average accretion rate ($\log \langle \lambda_\text{sBHAR} \rangle$) as presented by \cite{Aird18}. Specifically, they obtained $\lambda_\text{sBHAR}$ distributions in the field that peaks around $\log \langle \lambda_\text{sBHAR} \rangle \approx -1$ for star-forming galaxies at $z>2$ and $\log \langle \lambda_\text{sBHAR} \rangle \approx -0.5$ at $z>3$. Thus, the average $\lambda_\text{sBHAR}$ distribution of AGNs in protoclusters does not show significant deviation from the larger AGN sample in \cite{Aird18}. However, this comparison should be taken with caution as it could be affected by several factors, such as the broad distribution of $\lambda_\text{sBHAR}$ values observed, the limited sample size, AGN activity variability, the uncertainties in $\lambda_\text{sBHAR}$ measurements, and different methods in the selection of the galaxy populations, as well as the determination of $M_*$ and $\text{SFR}$.

The PDFs of $\lambda_\text{sBHAR}$, indicated by solid lines, reveal groups of AGNs with specific $\lambda_\text{sBHAR}$ values. In the MQN01 and Spiderweb protoclusters, some AGNs are found with $\lambda_\text{sBHAR}$ values near the Eddington limit. The presence of these AGN could indicate that, within the same overdensity, different galaxies could be subject to different environmental (e.g., varying amounts of surrounding gas and galaxies) and non-environmental (e.g., different internal gas transport mechanisms, accretion disc instabilities, feedback cycles) factors leading them to experience different accretion phases.
\begin{figure}[t]
   \begin{center}   
   \includegraphics[height=0.3\textheight,angle=0]{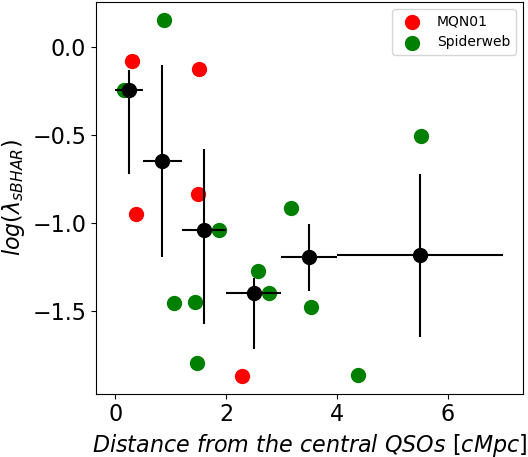}
   \caption{$\lambda_\text{sBHAR}$ as a function of the distance from the central QSOs, CTS G18.01 in MQN01 (red) and PSK1138 in the Spiderweb (green) protocluster. Black dots mark the median values of $\lambda_\text{sBHAR}$ for the AGNs in both protoclusters across different bins of distance from the two central QSOs. The uncertainties correspond to the 16th and 84th percentile values within each radial bin.}
   \label{fig:distLamb}
   \end{center}
\end{figure}
In Figure~\ref{fig:distLamb}, we explore whether such differences in $\lambda_\text{sBHAR}$ across AGN in MQN01 and Spiderweb could depend on the projected distance from the center of the protocluster (defined as the position of the central QSO in MQN01 and PKS 1138-262 in Spiderweb). 
The black dots represent the median $\lambda_\text{sBHAR}$ values, with error bars indicating the 16th and 84th percentiles, for AGNs in both protoclusters across different radial bins from the main QSOs.
Despite the limited sample size, both Pearson and Spearman correlation analyses indicate a strong and statistically significant negative relationship between $\lambda_\text{sBHAR}$ and the distance from the protocluster center in Spiderweb and MQN01 (i.e., black dots). The Pearson correlation shows a mean value of -0.71 with a 95$\%$ confidence interval of [-0.90, -0.52], while the Spearman correlation is slightly stronger, with a mean value of -0.76 and a 95$\%$ confidence interval of [-0.83, -0.60]. These results suggest that SMBH accretion rates are higher in the central, denser regions of protoclusters.

\section{Summary and Conclusions} \label{sec:sumconc}

We presented the analysis of deep ($\sim \rm 634~ks$) Chandra observations in a field known as MQN01 centered on a luminous QSO at $z \sim 3.2502$. The field contains a giant \lya\ nebula (\citealt{Borisova16}, \citepcantalupo) and one of the largest overdensities of UV continuum selected galaxies found so far at this redshift \citep[][\citepgalbiati]{Pensabene24}.
The deep Chandra observations have been used to obtain a comprehensive census of AGN embedded into this protocluster and in order to derive the hard X-ray luminosity function (HXLF), the overdensity ($\delta$), the fraction of AGN ($f_{AGN}$) and the accretion rates of the SMBHs ($\lambda _\text{sBHAR}$) in this field. These quantities have been compared to similar properties obtained for other two protoclusters available in the literature with deep Chandra observations, i.e., Spiderweb \citep{Tozzi22} and SSA22 \citep{Lehmer09}. 
The key findings are summarized below:
\begin{itemize}
	\item We have identified six X-ray AGN with $\rm \log(L_{2-10~keV}/erg~s^{-1}) \gtrsim 43.5$ among the spectroscopically selected protocluster members within an area of $\approx 16 \rm ~cMpc^2$ ($\rm \approx 2 \times 2 \rm~ arcmin^2$) and a velocity range of $\pm 1000~\rm km~s^{-1}$ around the most luminous QSO. The AGN overdensity above the survey luminosity within this volume, excluding the QSO, corresponds to $\langle \delta_{16} \rangle \simeq 10^3$ . This is one of the largest AGN overdensity found so far at $z>2$, exceeding, on average, the overdensities calculated within similar volumes and luminosities within the Spiderweb field ($\langle \delta_{16} \rangle \simeq 600$) and SSA22 ($\langle \delta_{16} \rangle \simeq 400$) as shown in Figure~\ref{fig:OverDen}. Moreover, the AGN HXLF in MQN01, similarly to the Spiderweb, are much flatter compared to those observed in the field at the same redshift for $\rm \log(L_{2-10~keV}/erg~s^{-1}) > 44.5$, showing an even higher overdensity of high-luminosity AGN compared to the field above these luminosities.
	\item The total fraction of AGN above the survey luminosity derived in MQN01 ($f_{AGN} > 0.2$) exceeds those calculated in low-redshift galaxy clusters and in the field ($f_{AGN} < 0.02$), and is comparable to those estimated in other protoclusters \citep{Tozzi22,Vito23}. The $f_{AGN}$ in the MQN01, similarly to the Spiderweb, increases substantially considering galaxies of higher stellar mass, as shown in Figure~\ref{figure:MSFR}. Surprisingly, the four most massive galaxies spectroscopically selected with MUSE in the MQN01 exceeding $\rm \log(M_*/M_{\odot}) > 10.5$ are all AGN. This is different than what observed in the Spiderweb field for which $f_{AGN}$ remains around or below 50$\%$ above these masses considering three different selection techniques. However, we caution that this result is sensitive to the low statistics and the particular selection technique in the MQN01 field. 
	\item  The $\lambda_\text{sBHAR}$ PDFs indicate that MQN01 hosts a higher fraction of rapidly accreting SMBHs compared to SSA22 and Spiderweb. Some AGNs in MQN01 and Spiderweb approach the Eddington limit. A wide distribution of $\lambda_\text{sBHAR}$ of about $2~\rm dex$ within the same overdensity suggests variations in galaxy accretion phases and surrounding gas availability can significantly influence SMBH growth. The $\lambda_\text{sBHAR}$ increases with proximity to the central QSOs in the MQN01 and Spiderweb protoclusters, suggesting indeed that denser environments could enhance SMBH accretion rates.
\end{itemize} 
In conclusion, our results suggest that protocluster environments such as the one probed by the MQN01 field hosts physical conditions that are ideal to promote the triggering and growth of X-ray AGN, especially for galaxies with stellar masses above $\log(M_*/\rm M_{\odot}) = 10.5$. This configuration represents a rare occurrence for an AGN population, and it is even more exceptional considering the variability in X-ray luminosity during the AGN phase. However, as suggested in the literature by simulations \citep[see][]{DiMatteo05b,Jiang24}, the high-density regions, especially those with large gas reservoirs at these redshifts \citep[e.g.,][]{Zou24}, may play a crucial role in prolonging the period during which a SMBH can sustain rapid accretion. 
In the second part of this project, we will focus on the individual properties of AGN host galaxies within the MQN01 protocluster, examining them within the broader context of co-evolving SMBHs and large-scale structures.

In the analyses presented in this paper, Poissonian uncertainties provide a challenge for the determination of the role of environmental factors and galaxy properties on SMBH accretion and growth. To draw more definitive conclusions, future studies will need to perform homogeneous observations across multiple wavelengths in diverse protocluster fields, covering a broad range of redshifts and overdensities. 
Therefore, further deep X-ray surveys of protoclusters are essential to expand these studies. X-ray missions with greatly increased sensitivity and field of view, such as Athena \citep{Nandra13} and Lynx \citep{Gaskin19}, would overcome the limited statistics and allow the extension of these studies to many more protoclusters and to higher redshifts.

\begin{acknowledgements}
This project was supported by the European Research Council (ERC) Consolidator Grant 864361 (CosmicWeb). FF acknowledges supported by HORIZON2020: AHEAD2020-Grant Agreement n. 871158.
\end{acknowledgements}

\bibliographystyle{aa} 
\bibliography{andrea}


\appendix

\section{Zoom-in of the count maps of the X-ray Protocluster Members and their Spectra} \label{app:zoomin}

\begin{figure*}[t]
   \begin{center}
   \includegraphics[height=0.478\textheight,angle=0]{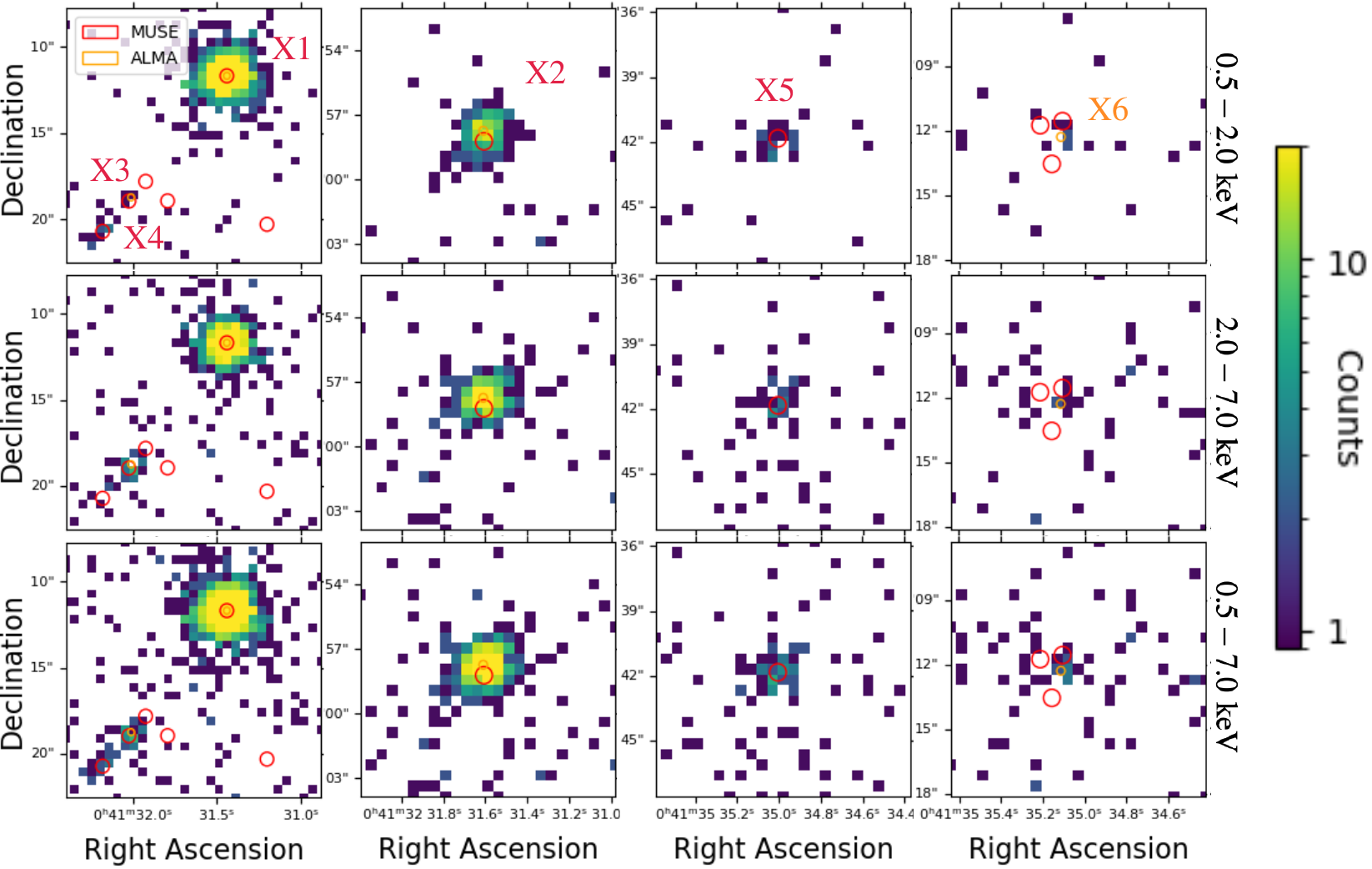}
   \caption{Chandra X-ray images (in counts) focusing on the six X-ray sources associated with spectroscopically-selected protocluster members. The images  are binned in energy bands: soft (0.5-2.0 keV) in the top panels; hard (2.0-7.0 keV) in the central panels; and broad (0.5-7.0 keV) in the bottom panels. Red and orange circles denote the protocluster members identified through spectroscopic selection with MUSE and ALMA, respectively. The first column depicts the region of the three closely located X-ray detections within a FoV of $115 \times 115 ~\rm kpc^2$, while the subsequent columns present FoVs of approximately $92 \times 92 ~\rm kpc^2$, centered on the other detections.}
   \label{fig:zoomcounts}
   \end{center}
\end{figure*} 

In this section we report the Chandra X-ray images in the soft (0.5-2.0 keV; top panel), hard (2.0-7.0 keV; central panel), and broad (0.5-7.0 keV; bottom panel) energy bands, zoomed in on the six X-ray sources associated with spectroscopically-selected protocluster members, in Figure~\ref{fig:zoomcounts}. This image is essential for highlighting the significant number of counts in each band for the X-ray sources associated with the protocluster members. Moreover, it illustrates our ability to deblend the X-ray detections of ID3 and ID4. Specifically, ID3 is primarily visible in the hard X-ray band, while ID4 is more prominent in the soft band. This allows us to confidently determine that these two slightly blended sources are associated to two distinct MUSE-selected protocluster members.

\begin{figure*}[t]
   \begin{center}   
   \includegraphics[height=0.8\textheight,angle=0]{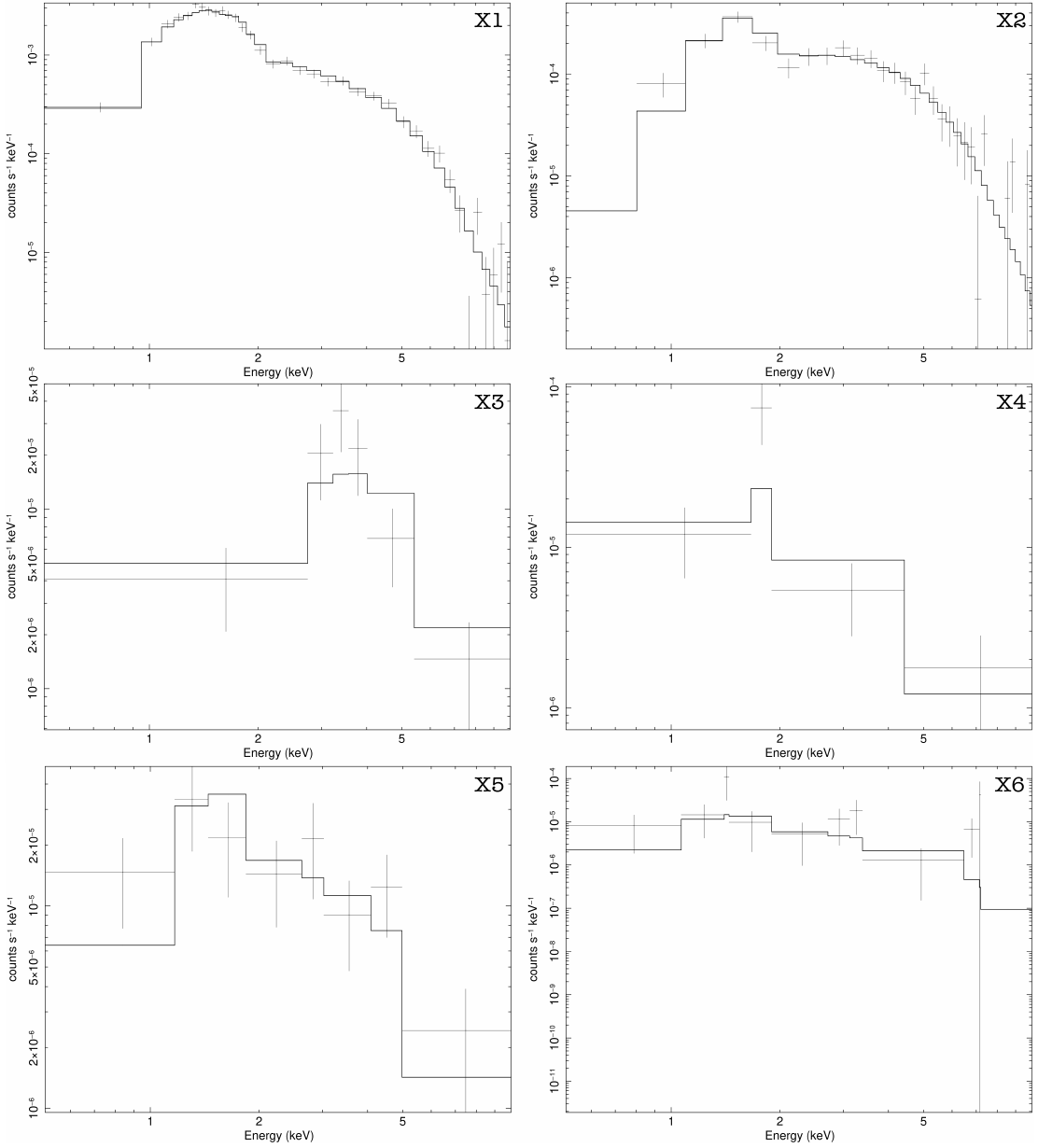}
   \caption{Spectra and best-fit models of the six X-ray AGNs embedded into the MQN01 protocluster. Spectra are rebinned to ensure a minimum of 20 counts per energy bin (top panels) and 3 counts per energy bin (middle and bottom panels). See Section~\ref{sec:spectraanalysis} for further details.}
   \label{fig:xspectra}
   \end{center}
\end{figure*}

In Figure~\ref{fig:xspectra}, we present the spectra of the X-ray sources in the MQN01 protocluster along with their best-fit spectral models. Spectra from X-ray sources with more than 50 net counts in the broad (0.5-7 keV) band have been rebinned to ensure a minimum of 20 counts per energy bin, while spectra with fewer counts have been rebinned to achieve a minimum of 3 counts per energy bin. Details of the spectral fitting procedures are provided in Section~\ref{sec:spectraanalysis}.

\end{document}